\documentclass{article}

\usepackage{amsfonts}
\usepackage{amsmath}
\usepackage{amsthm}
\usepackage{arxiv}
\usepackage{booktabs}
\usepackage{doi}
\usepackage[mathscr]{eucal}
\usepackage[T1]{fontenc}
\usepackage{graphicx}
\usepackage{hyperref}
\usepackage[utf8]{inputenc}
\usepackage{mathrsfs}
\usepackage{microtype}
\usepackage{nicefrac}
\usepackage{url}
\usepackage{xcolor}

\numberwithin{equation}{section}

\theoremstyle{plain}

\newtheorem{theorem}{Theorem}
\newtheorem{corollary}{Corollary}

\theoremstyle{definition}

\theoremstyle{remark}

\title{The Instability of the Critical Friedmann Spacetime at the Big Bang as an Alternative to Dark Energy}

\author{ \href{https://orcid.org/0000-0001-9255-6281}{\includegraphics[scale=0.06]{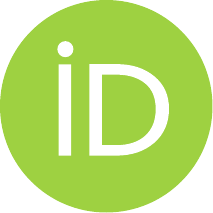}\hspace{1mm}Christopher Alexander}\\
Department of Mathematics\\
University College London\\
London, WC1H 0AY\\
United Kingdom\\
\texttt{christopher.alexander@ucl.ac.uk}\\
\And
\href{https://orcid.org/0000-0002-6907-1101}{\includegraphics[scale=0.06]{orcid.pdf}\hspace{1mm}Blake Temple} \\
Department of Mathematics\\
University of California\\
Davis, CA 95616\\
United States\\
\texttt{temple@math.ucdavis.edu}\\
\And
{Zeke Vogler} \\
Department of Mathematics\\
University of California\\
Davis, CA 95616\\
United States\\
\texttt{zekius@math.ucdavis.edu}
}

\renewcommand{\headeright}{}
\renewcommand{\undertitle}{}
\renewcommand{\shorttitle}{}

\hypersetup{
	pdftitle={The Instability of the Critical Friedmann Spacetime at the Big Bang as an Alternative to Dark Energy},
	pdfsubject={gr-qc, math-ph},
	pdfauthor={Christopher Alexander},
	pdfkeywords={General Relativity, Instability, Cosmology, Dark Energy},
}

\begin{document}

\maketitle

\begin{abstract}

We characterize the local instability of pressureless Friedmann spacetimes to radial perturbation at the Big Bang. The analysis is based on a formulation of the Einstein--Euler equations in self-similar variables $(t,\xi)$, with $\xi=r/t$, conceived to realize the critical ($k=0$) Friedmann spacetime as a stationary solution whose character as an unstable saddle rest point $SM$ is determined via an expansion of smooth solutions in even powers of $\xi$. The eigenvalues of $SM$ imply the $k\neq0$ Friedmann spacetimes are unstable solutions within the unstable manifold of $SM$. We prove that all solutions smooth at the center of symmetry agree with a Friedmann spacetime at leading order in $\xi$, and with an eye toward Cosmology, we focus on $\mathcal{F}$, the set of solutions which agree with a $k<0$ Friedmann spacetime at leading order, providing the maximal family into which generic underdense radial perturbations of the unstable critical Friedmann spacetime will evolve. We prove solutions in $\mathcal{F}$ generically accelerate away from Friedmann spacetimes at intermediate times but decay back to the same leading order Friedmann spacetime asymptotically as $t\to\infty$. Thus instabilities inherent in the Einstein--Euler equations provide a natural mechanism for an accelerated expansion without recourse to a cosmological constant or dark energy.

\end{abstract}


C.A. gratefully acknowledges the support of the EPSRC project EP/S02218X/1 and the ERC starting grant 101078061 SINGinGR, under the European Union's Horizon Europe program for research and innovation.

This research was also supported by the American Institute of Mathematics SQuaRE Program.

\vfill

\pagebreak

\tableofcontents

\vfill

\pagebreak

\section{Introduction}\label{S1}

The Friedmann\footnote{By Friedmann spacetimes, we mean the family of Friedmann--Lemaître--Robertson--Walker (FLRW) spacetimes parameterized by curvature constant $k\in\mathbb{R}$ \cite{ABS1975}.} family of spacetimes has been the starting point for modern Cosmology since Lemaître and Hubble first formulated the theory of an expanding universe of galaxies emanating from an initial big bang singularity. The theory is based on the explicit solutions of Einstein's field equations, discovered by Alexander Friedmann in the early 1920s. As the story goes, in 1922 Friedmann, working alone in St. Petersburg, Russia, sent his solutions to Einstein, who initially rejected them believing the Universe was static but shortly thereafter accepted them as correct under appeal by Friedmann. Friedmann died soon after and the physical interpretation of his solutions was taken up by Eddington, Lemaître and Hubble in the late 1920s. By 1931, Einstein rejected the static model as unstable and accepting Hubble's 1929 measurement of the expanding universe, is famously quoted as describing Lemaître's Cosmology based on Friedmann spacetimes as \emph{the greatest theory of creation in history}. In the present paper, we give a theorem stating that the Friedmann spacetimes are in fact all unstable to radial perturbation, at every order.

The Friedmann family of spacetimes describe a uniform three-dimensional universe of galaxies expanding in time from an initial big bang singularity at a rate determined by Einstein's field equations. The Friedmann spacetimes derived in comoving coordinates contain a single real valued parameter $k$, which determines the sign of the scalar curvature of an expanding three-dimensional space of constant density and curvature at each fixed time.\footnote{The curvature parameter $k$ can always be rescaled to one of the values $k=-1,0,1$. The $k=0$ Friedmann spacetime is seen to be unique and the $k=\pm1$ spacetimes each describe a one parameter family of distinct spacetimes depending on the single parameter $\Delta_0$ \cite{ATV2024}. Thus, for example, we refer to the one parameter family of underdense $k<0$ Friedmann spacetimes by $k$ or $\Delta_0$. Unless a different equation of state is specified, our use of the term Friedmann always assumes a zero pressure (dust) solution of the Einstein--Euler equations, typically taken as a good physical approximation after the pressure drops precipitously to zero an order of magnitude in time before the uncoupling of matter and radiation \cite{P1999}.} In the early 1930s, Robertson and Walker anchored the acceptance of Cosmology based on Friedmann spacetimes with the so called Copernican Principle, by establishing mathematically that if there is no special place in a three-dimensional universe of galaxies, made precise by the condition that space be homogeneous and isotropic about every point, then any such three-space evolving in time according to Einstein's field equations must indeed be a Friedmann spacetime. This connected the Friedmann spacetimes with prescientific notions of Earth not being in a special place in the Universe, a sort of principle of Physics more or less accepted as a justification in the community of established cosmologists since that time. The high degree of uniformity in the microwave background radiation and distribution of the galaxies lend strong support to the Copernican Principle, supporting the starting assumption that the Universe of galaxies is, on the largest scale, a Friedmann spacetime, implying the existence of a big bang singularity at some initial time. Fluctuations in the microwave background radiation argue strongly for the critical ($k=0$) Friedmann spacetime, the unique self-similar member of the Friedmann family in which the expanding three-space at each fixed time is perfectly flat.

A major challenge to this theory arose in 1999 with the discovery of the anomalous acceleration of the galaxies based on supernova data. To maintain the critical Friedmann solution and consistency with both the Einstein field equations and the observed expansion rate of the galaxies based on that data, the cosmological constant was introduced back into Einstein's equations for late time Cosmology, interpreted as dark energy.\footnote{Einstein introduced the cosmological constant initially to make his equations consistent with a static universe but this turned out to be unstable, so he took it back in the early 1930s, famously quoted as saying the cosmological constant was the greatest blunder of his career, as he could have predicted the expanding universe confirmed by Hubble. Solutions with pure cosmological constant are currently the basis for theories of inflation, which are not considered here \cite{BGG1987,GH2007,P1999}.} Using the supernova data, the best fit among Friedmann spacetimes with dark energy was deduced to be the critical Friedmann spacetime with a cosmological constant. To fit the supernova data with the critical Friedmann spacetime requires approximately $70\%$ of the energy density of the Universe at present time to consist of anti-gravitating dark energy \cite{ABNV2009,CFL2008,CF2009,KMR2006,L+2025,P2003,P+1999,R+1998,R2010,STV2017,ST2012A,TS2009,V2006}. This is the basis for the current $\Lambda CDM$ model of Cosmology, which appears to account for most of the cosmological data.

One worrisome problem with the $\Lambda CDM$ model, aside from the fact that dark energy is a mysterious anti-gravitating source of energy not known to connect with any other physical theory \cite{GH2007,P1999,ST2012A,TS2009,W1972},\footnote{A nonzero cosmological constant is not needed in physical confirmations of Einstein's theory other than for Cosmology. Furthermore, since the energy density decreases like $t^{-2}$ and $\Lambda$ is constant, the $\Lambda CDM$ model also places the Universe in a \emph{special place in time} when the energy density of classical matter and dark energy are on the same order, a different seeming violation of the Copernican Principle and the hallmark of using an ad hoc correction to the equations when its the underlying solution that's wrong \cite{ST2009,ST2012B}.} is that the Friedmann spacetimes are unstable to radial perturbation. Indeed, physically, galaxy formation is presumed to have occurred due to local gravitational instabilities \cite{L2008}, and it has been known as well that the critical Friedmann spacetime is formally unstable to (nonlocal) perturbation in the curvature parameter $k$ \cite{GH2007,STV2017} (see also Section \ref{S1.3}).

In this paper we give what, to authors' knowledge, is the first definitive \emph{local} characterization of the instability to radial perturbation present in the critical and non-critical Friedmann spacetimes at (and hence near) the Big Bang. The analysis is based on a derivation of the Einstein--Euler equations for smooth spherically symmetric spacetimes in standard Schwarzschild coordinates (SSC), expressed in self-similar variables $(t,\xi)$, with $\xi=r/t$, conceived in \cite{STV2017} to realize the critical Friedmann spacetime as a stationary solution. Here we complete the analysis by determining the character of the critical Friedmann spacetime as an unstable saddle rest point $SM$ via a definitive eigenvalue analysis at every order of an expansion of smooth solutions in even powers of $\xi$. We then prove that non-critical Friedmann spacetimes lie in the unstable manifold of $SM$ at every order, which characterizes their instability according to the positive eigenvalues of $SM$. In particular, since the critical Friedmann spacetime is self-similar (see \cite{CT1971}), this implies that all Friedmann spacetimes approach self-similarity at the Big Bang ($t=0$) at every order. By introduction of a solution dependent time translation which we call \emph{time since the Big Bang}, we prove that every smooth solution of our self-similar equations agrees exactly with a Friedmann spacetime at leading order in $\xi$, but generic solutions outside the unstable manifold of $SM$ do not exhibit a self-similar big bang like Friedmann spacetimes do above leading order.\footnote{This is consistent with the self-similarity hypothesis \cite{CC2005}.}

With potential cosmological implications in mind, we use the local stability results to obtain a \emph{global} characterization of accelerations away from $k<0$ Friedmann spacetimes induced by underdense perturbations of the critical ($k=0$) Friedmann spacetime at $t=0$, the choice $k<0$ being consistent with the sign of $\Lambda$ in the theory of dark energy. For this, we identify the new family $\mathcal{F}$ of solutions underdense with respect to the $k=0$ Friedmann spacetime, defined as the set of all solutions which agree with a $k<0$ Friedmann spacetime at leading order in $\xi$. We prove that solutions in $\mathcal{F}$, both inside and outside the unstable manifold of $SM$, ubiquitously accelerate away from Friedmann spacetimes at intermediate times but decay back to the same leading order Friedmann spacetime as $t\to\infty$.\footnote{For each fixed $\bar{r}>0$, see Theorem \ref{T1}.} This shows that instabilities inherent in the Einstein--Euler equations naturally induce accelerations away from Friedmann spacetimes.

We propose $\mathcal{F}$ as the maximal asymptotically stable family of solutions into which generic underdense radial perturbations of the unstable $k=0$ Friedmann spacetime will evolve and naturally admit accelerations away from Friedmann spacetimes within the dynamics of Einstein's original equations, that is, without recourse to a cosmological constant or dark energy. The resulting definitive stability analysis of Friedmann spacetimes is of mathematical and physical interest in its own right but we present this with an eye toward exploring the possibility that accelerations away from Friedmann spacetimes generated by the instability at the Big Bang could potentially account for the anomalous acceleration of the galaxies. Indeed, every solution in $\mathcal{F}$, defined after some initial time $0<t_0<<1$, evolves from a big bang associated with the Radiation Dominated Epoch (RDE) by simply imposing boundary conditions at $t=t_0$ (as done in \cite{STV2017}) and running the resulting RDE solution backwards into a singularity.\footnote{Accelerations over and above Friedmann spacetimes have a center of expansion and this has historically been viewed as a violation of the Copernican Principle. Note there is a small angular dependence in the microwave background radiation \cite{CHSS2006} and all current models seem to place Earth in some sort of \emph{special place}, suggesting to the authors that some violation of the Copernican Principle might be something we are forced to accept. Previous efforts to model the anomalous acceleration as a local underdensity that tends to the $k=0$ Friedmann spacetime in the far field based on comoving Tolman--Bondi coordinates were criticized in \cite{GH2008,GH2007,P1999}. The analysis here appears to support these criticisms, as it is difficult to see how such a regular matching would be viable given that our analysis in self-similar variables shows that all underdense solutions evolve to a $k<0$ Friedmann spacetime in the far field, a consequence of our proof that all underdense solutions decay to a rest point $M$ as $t\to\infty$.}

Authors' program to address these issues began in our 2017 announcement in RSPA \cite{STV2017}. Unfortunately, our collaborator Joel Smoller died shortly before publication.\footnote{Authors dedicate this paper to our former collaborator and long-time friend Joel Smoller and acknowledge the use of unpublished notes associated with publication \cite{STV2017}.} In \cite{STV2017} we introduced the self-similar version of the Einstein--Euler equations employed here, together with an expansion of the equations in even powers of $\xi$. This was for the purpose of evolving a self-similar underdense wave at the end of the RDE, up to present time, to make a prediction of the resulting acceleration in comparison to the acceleration due to dark energy. In this new announcement we complete the mathematical analysis of the instability of $p=0$ Friedmann spacetimes begun in \cite{STV2017} via a definitive eigenvalue analysis of $SM$ sufficient to incorporate $k\neq0$ Friedmann spacetimes. From this, a new unifying picture of the instability of the entire family of Friedmann spacetimes emerges with interesting surprises in both the theory and its implications. One difference in point of view here is that the corrections to the $k=0$ Friedmann spacetime were viewed in \cite{STV2017} as coming from fluctuations associated with an earlier epoch, while here corrections to the $k=0$ Friedmann spacetime emerge directly from the unstable manifold of the $k=0$ Friedmann spacetime at the Big Bang itself. That is, $k<0$ Friedmann spacetimes emerge at the Big Bang as particular solutions within the underdense component $\mathcal{F}$ of the unstable manifold associated with the $k=0$ Friedmann spacetime, represented as a saddle rest point $SM$ in self-similar variables $(t,\xi)$, while every other solution in $\mathcal{F}$ exhibits accelerations away from $k<0$ Friedmann spacetimes. To prove that $k\neq0$ Friedmann spacetimes lie in $\mathcal{F}'$, the subset of $\mathcal{F}$ which lie in the unstable manifold of $SM$, we prove it for the ODE at order $n=2$ and then deduce $n>2$ from the fact that all eigenvalues of $SM$ above order $n=2$ are positive. The existence of a second positive eigenvalue at order $n=2$, distinct from the leading order positive eigenvalue associated with Friedmann spacetimes, then establishes the instability at the Big Bang of $k\neq0$ Friedmann spacetimes within the unstable manifold of $SM$ at order $n=2$, with this instability appearing at every order of our expansion. The existence of a highest order negative eigenvalue at order $n=2$ establishes that solutions within the unstable manifold of $SM$ are non-generic.

With regard to the global dynamics, we prove that all solutions in $\mathcal{F}$ are asymptotically stable in the sense that, as $t\to\infty$, all solutions in $\mathcal{F}$ follow $\mathcal{F}'$ into a universal stable rest point $M$ (for Minkowski) at every order of expansion in even powers of $\xi$. Solutions in $\mathcal{F}$ align exactly with a $k<0$ Friedmann spacetime at leading order in $\xi$, generically introduce higher order accelerations away from $k<0$ Friedmann spacetimes at intermediate times, then decay back to the same $k<0$ Friedmann spacetime uniformly at each fixed radius $r>0$ as $t\to\infty$. The analysis establishes that the locally spatially homogeneous self-similar Big Bang of Friedmann spacetimes is universal among solutions in $\mathcal{F}'$ and in all of $\mathcal{F}$ at leading order in $\xi$, but surprisingly, the existence of a negative eigenvalue at $SM$ at order $n=2$ implies that generically in $\mathcal{F}$ the Big Bang is not self-similar as $t\to0$ beyond leading order.

In the remainder of the introduction we describe our results in detail. Rigorous proofs are accomplished in \cite{ATV2024}.

\subsection{Mathematical Background}\label{S1.1}

In our 2017 announcement in RSPA, Smoller, Temple and Vogler introduced the STV PDE, a version of the Einstein--Euler equations for spherically symmetric self-similar spacetimes. These were obtained by starting with a spacetime metric in standard Schwarzschild coordinates (SSC) and re-expressing it in self-similar variables, that is, coordinates which employ the self-similar variable $\xi=r/t$ as the spatial coordinate in place of the SSC radial coordinate $r$, assuming zero cosmological constant and $0<\xi<1$ to keep $\xi$ as a spacelike coordinate \cite{STV2017}. Since $\xi=1$ is a measure of the distance of light travel since the Big Bang in a Friedmann spacetime, we view $0<\xi<1$ as a description of radially symmetric cosmological models valid out to approximately the Hubble radius, a measure of the distance across the Visible Universe \cite{ST2012A,TS2009}. The STV PDE were conceived to represent the pressureless ($p=0$) critical ($k=0$) Friedmann spacetime as a stationary solution which we denote by $SM$ for \emph{Standard Model}.\footnote{A \emph{stationary solution} or \emph{rest point} of the STV PDE is a solution which depends only on $\xi$, and hence is independent of time $t$ when $\xi$ is taken to be the spatial coordinate.} This was motivated by the important observation that the metric components and fluid variables of the $p=0$, $k=0$ Friedmann spacetime in SSC can be expressed as a function of $\xi$ alone when the time gauge is taken to be geodesic (proper) time since the Big Bang at the center of expansion $r=\xi=0$ (see \cite{STV2017} and Theorem \ref{T1}). In this paper we extend the derivation of the STV PDE to the case $p=\sigma\rho$, with $\sigma$ constant, which incorporates both the pressureless ($\sigma=0$) and pure radiation ($\sigma=\frac{1}{3}$) cases and represents the $k=0$ Friedmann spacetime as a rest point of the equations for every $0\leq\sigma<1$. In the $p=0$ case the rest point is isolated, in the $p\neq0$ case the rest point is not isolated due to the existence of a one parameter family of self-similar perturbations of the $k=0$ Friedmann spacetime \cite{A2021,ST2012A,TS2009}. In this paper we restrict attention to analyzing perturbations in the $p=0$ case.

The character of a rest point of an ODE or PDE is difficult to disentangle in coordinate systems where it appears time dependent, which is the case for $SM$ in comoving coordinates \cite{STV2017}. To analyze rest points of a PDE requires a procedure of finite approximation and this was manifested in \cite{STV2017} by the observation that solutions of the STV PDE which are smooth at the center of symmetry can be developed into a regular expansion in even powers of $\xi$ with time dependent coefficients. This generates a sequence of autonomous $2n\times2n$ ODE which close in the time dependent coefficients at every order $n\geq1$. We name these the STV ODE of order $n$. The STV ODE are nested in the sense that higher order solutions provide strict refinements of solutions determined at lower orders. Each new order $n\geq1$ introduces two new variables, a density variable $z_{2n}$ and a velocity variable $w_{2n-2}$ (defined in Section \ref{S2.3}), whose time dependence is determined from two initial conditions free to be imposed at each order $n\geq1$.\footnote{This asymptotic expansion does not close at order $n$ when $p\neq0$ \cite{STV2017}.}

\subsection{The STV ODE}\label{S1.2}

We establish here that the STV ODE at each order are autonomous in the log-time variable $\tau=\ln t$, and the phase portrait at each order contains the unstable saddle rest point $SM$ together with the stable degenerate rest point $M$. Being autonomous, each trajectory in the phase portrait of an STV ODE represents a one parameter family of distinct solutions determined by translation in $\tau$ \cite{S1994}. The analysis in \cite{STV2017} only required that the STV ODE be constructed up to orders $n=1$ and $n=2$ and the instability of the $k=0$ Friedmann spacetime was described at these orders. Here we derive a closed form expression for the STV ODE at all orders, determine the eigenvalues of rest points $SM$ and $M$ at every order and incorporate the $k\neq0$ Friedmann spacetimes into the stability analysis by use of known exact formulas. For the analysis we introduce a new solution dependent time gauge for SSC spacetimes which we call \emph{time since the Big Bang}, the purpose of which is to normalize general solutions of the STV ODE and STV PDE relative to Friedmann spacetimes. Employing time since the Big Bang, we prove that every solution of the STV ODE and STV PDE is gauge equivalent to a solution which agrees exactly with a Friedmann spacetime at leading order $n=1$ but solutions generically diverge from Friedmann spacetimes at higher orders. We prove that the rest point $M$ is the time asymptotic limit $t\to\infty$ of solutions underdense relative to the $k=0$ Friedmann spacetime, where we define a solution to be \emph{underdense} if and only if it agrees with an underdense ($k<0$) Friedmann spacetime at leading order alone. Using this we characterize the phase portrait of \emph{all} underdense solutions of the STV ODE at every order and this provides a \emph{global} characterization of the instability of $k\leq0$ Friedmann spacetimes to underdense perturbations. The presence of the rest point $SM$, together with the rest point $M$ which characterizes the late time dynamics of all underdense solutions at every order of the STV ODE, appears to be a serendipitous simplification inherent in the choice of self-similar standard Schwarzschild coordinates.

We prove here that when the lower order terms are viewed as known (consistent with the nested property of the STV ODE), the STV ODE are linear inhomogeneous equations at every order $n\geq2$. Thus, being linear, solutions of the STV ODE exist for all time, for arbitrary initial data, at every order $n\geq1$. Our analysis shows that the eigenvalue structures of the rest points $SM$ and $M$ determine the character of the phase portrait of the STV ODE at every order, and because the eigenvalues of $SM$ at order $n\geq3$ are all distinct and positive, the essential character of all higher order phase portraits can be deduced from the phase portraits at orders $n=1$ and $n=2$. By use of exact formulas, we establish that $k<0$ and $k>0$ Friedmann spacetimes correspond to unique opposite solution trajectories which lie in the unstable manifold of $SM$ at all orders of the STV ODE when time since the Big Bang is imposed. Imposing the solution dependent time since the Big Bang gauge has the effect of eliminating the leading order negative eigenvalue from the analysis by transforming each solution of the STV ODE to a gauge equivalent solution which agrees with $SM$, or else with one of the two branches of the unstable manifold of $SM$ at order $n=1$.\footnote{This point was not understood in \cite{STV2017}.} In other words, the $k<0$ Friedmann spacetimes extend to a general family of solutions defined to be underdense relative to the $k=0$ Friedmann spacetime by simply defining the \emph{underdense} solutions of the STV ODE or STV PDE to be those which project to a solution on the portion of the unstable manifold of $SM$ associated with $k<0$ Friedmann spacetimes at order $n=1$ (the trajectory which connects $SM$ to $M$ in Figure \ref{Figure2}) when the time gauge is taken to be time since the Big Bang. Thus we identify a new family of solutions $\mathcal{F}$ defined to be the collection of all solutions of the STV PDE which are underdense relative to the $k=0$ Friedmann spacetime at order $n=1$. The main topic of this paper is to characterize the local instability of Friedmann spacetimes at $t=0$ and describe the global instability of $k<0$ Friedmann spacetimes via a characterization of solutions in $\mathcal{F}$.

\subsection{The Instability of Friedmann Spacetimes}\label{S1.3}
 
The instability of the critical ($k=0$) Friedmann spacetime follows from the structure of the eigenvalues at $SM$, and we prove that the instability of non-critical ($k\neq0$) Friedmann spacetimes follows from the structure of the unstable manifold of $SM$. Regarding the eigenvalues at $SM$, we show that two new and distinct eigenvalues of $SM$ emerge at every order, specifically, one positive and one negative eigenvalue at orders $n=1$ and $n=2$, and all positive eigenvalues at orders $n\geq3$. Imposing time since the Big Bang eliminates the leading order negative eigenvalue by transforming every solution at order $n=1$ to either $SM$ or to one of the two branches of the unstable manifold of $SM$. This eliminates one dimension of possible trajectories at every order, leaving $2n-1$ degrees of freedom in the solutions, that is, there are $2n-2$ degrees of freedom to choose the trajectory and one remaining degree of freedom to choose log-time translation along each trajectory in each $2n\times2n$ STV ODE of order $n$. The local instability of the $k=0$ Friedmann spacetime is determined at order $n=1$ by the $k\neq0$ Friedmann spacetimes alone, because all solutions agree with a Friedmann spacetime at order $n=1$. Thus order $n=2$ is the lowest order at which the instability of the $k\neq0$ Friedmann spacetimes is triggered and this is characterized in the phase portrait of the $4\times4$ STV ODE of order $n=2$ as follows: At $n=2$ the $k=0$ Friedmann spacetime represented by $SM$ is an unstable saddle rest point with a codimension one (two-dimensional) unstable manifold of solutions and the $k\neq0$ Friedmann spacetimes are represented by unique and opposite solution trajectories which make up only one dimension of solutions within the two-dimensional unstable manifold of $SM$. Thus $k\neq0$ Friedmann spacetimes are locally unstable to a codimension one (one-dimensional) subspace of perturbations within the two-dimensional unstable manifold of trajectories associated with $SM$. Both statements remain qualitatively true at all orders $n\geq3$ as well, because the only negative eigenvalue at $SM$ appearing above order $n=1$ emerges at order $n=2$. It follows that a smooth solution of the STV PDE will lie in the unstable manifold of $SM$ at all orders $n\geq1$ of the STV ODE if and only if it lies in the unstable manifold of $SM$ at order $n=2$. From this we prove that when time since the Big Bang is imposed, the $k\neq0$ Friedmann spacetimes correspond to opposite sides of a single trajectory which lies in the unstable manifold of $SM$ at every order $n\geq1$ and at each order this trajectory is unstable to perturbation within a codimension one ($(2n-3)$-dimensional) space of solutions within the unstable manifold of $SM$, with the unstable manifold of $SM$ being a codimension one set of $2n-2$ trajectories within the full $2n-1$ space of independent solutions at order $n$. To prove solution trajectories of $k\neq0$ Friedmann spacetimes lie on unique trajectories in the unstable manifold of $SM$ at every order, we use exact formulas to establish this up to order $n=2$ and then apply our general result that a trajectory lies in the unstable manifold of $SM$ at every order if and only if it lies in the unstable manifold of $SM$ order $n=2$. This proves that $k\neq0$ Friedmann spacetimes are locally unstable at $t=0$ at every order $n\geq2$ of the STV ODE.

Regarding the global asymptotics of solutions, we prove that all underdense solutions of the STV ODE of order $n\geq1$, that is, all solutions in $\mathcal{F}$, are globally asymptotically stable in the sense that every solution in $\mathcal{F}$ converges to the unique rest point $M$ as $t\to\infty$ at every order $n\geq1$, meaning convergence to $M$ at order $n=1$ forces convergence to $M$ at every higher order as well. Moreover, at every order, each such underdense solution converges as $t\to\infty$\footnote{For each fixed $\bar{r}>0$, see Theorem \ref{T7}.} to the same $k<0$ Friedmann spacetime it agrees with at order $n=1$ and convergence to this Friedmann spacetime is faster than convergence to the rest point $M$ (precise decay rates are given in Section \ref{S2.6}). We deduce from this a simple characterization of the instability of $k<0$ Friedmann spacetimes at every order of the STV ODE: Underdense perturbations of a $k<0$ Friedmann spacetime generically accelerate away from Friedmann spacetimes at early and intermediate times until large time asymptotic decay to rest point $M$ brings the perturbation back to the unique $k<0$ Friedmann spacetime it agrees with at order $n=1$.

The existence of a negative eigenvalue at order $n=2$, and distinct eigenvalues at all higher orders, implies that, when time since the Big Bang is imposed, $SM$ is a non-degenerate saddle rest point with a one-dimensional stable manifold and codimension one unstable manifold in a $(2n-1)$-dimensional space of trajectories at every order $n\geq2$. An interesting consequence of the unique negative eigenvalue at order $n=2$ is that the backward time dynamics of solutions in $\mathcal{F}$ are generically not $SM$ except at leading order, so the Big Bang is always self-similar like the Big Bang of Friedmann spacetimes at order $n=1$ but is generically not self-similar at orders $n\geq2$.

The $k>0$ Friedmann spacetimes are similarly locally unstable at $t=0$ but they globally expand until the velocity reaches zero, at which point the solution evolves by time reversal. The global dynamics in the case $k>0$ will be considered in a subsequent paper containing details of the technical results. Regarding the global instability of Friedmann spacetimes, we focus here on the underdense case of solutions in $\mathcal{F}$ which agree with a $k<0$ Friedmann spacetime at order $n=1$, as this is most relevant to Cosmology, since $k<0$ is the case analogous with a positive cosmological constant, the case which introduces outward accelerations relative to the Standard Model of Cosmology without a cosmological constant.

\subsection{Time Since the Big Bang}\label{S1.4}

In \cite{STV2017}, we derived the $p=0$ STV PDE to describe the evolution of spacetimes in self-similar standard Schwarzschild coordinates under the assumption that the metrics are smooth and time is set to geodesic time at the center of symmetry. The time dependent coefficients of the Taylor expansion of such solutions in even powers of $\xi$ up to order $2n$ solve the STV ODE of order $n$ in variables $(z_{2n},w_{2n-2})$, and these variables determine all other unknowns in a solution (see Section \ref{S2.3}). However, there exists one final gauge freedom not fixed in the STV PDE and that is the time translation freedom of the SSC metric ansatz at the center $r=\xi=0$, a trivial gauge transformation of the SSC ansatz but non-trivial in self-similar coordinates because translating $t$ alters the physical position of the coordinate singularity in $\xi=r/t$. It is not possible to set this time gauge apriori without knowing the solution dependent time at which a spacetime metric exhibits an initial big bang type singularity, so it is incorporated into the STV PDE and STV ODE as a redundancy in physical solutions which presents itself as a negative eigenvalue at order $n=1$. In the present paper we prove that there exists a solution dependent SSC time translation $t\to t-t_0$ which we name \emph{time since the Big Bang}, such that making the SSC gauge transformation to time since the Big Bang has the effect of eliminating the leading order negative eigenvalue at $SM$. This is because it transforms each solution of the STV PDE to either the rest point $SM$ or one of the two trajectories in the unstable manifold of $SM$ at order $n=1$. As a result of this, every solution agrees exactly with a $k<0$, $k=0$ or $k>0$ Friedmann spacetime in the phase portrait of the STV ODE at leading order $n=1$. In other words, the existence of the time since the Big Bang gauge transformation establishes that the Big Bang singularity in each solution of the STV PDE has precisely the character of a Friedmann big bang singularity at leading order, for one of the three values $k=-1,0,1$.

There is an important distinction to make here: The STV ODE are autonomous in log-time $\tau=\ln t$, so translation in $\tau$ maps solutions to physically different solutions which traverse the same trajectory of the STV ODE, but translation in $t$ is a gauge freedom of the SSC metric ansatz which maps trajectories of the STV ODE to different trajectories which represent the same physical solution. Thus choosing time since the Big Bang eliminates a physical redundancy in the solution trajectories of the STV ODE and STV PDE. When time since the Big Bang is imposed, there are only three remaining trajectories in the leading order $n=1$ phase portrait of the STV ODE: The unstable rest point $SM$, the underdense (left) component of the unstable manifold and the overdense (right) component of the unstable manifold, as diagrammed in Figure \ref{Figure2}. For example, the underdense component of the unstable manifold of $SM$ at order $n=1$ is the unique trajectory which takes $SM$ to $M$ and coincides with $k<0$ Friedmann spacetimes at leading order, with the value of $k$ determined by translation in $\tau\to\tau-\Delta_0$ along this unique trajectory when $t$ is time since the Big Bang and $\Delta_0$ is the usual free parameter associated with Friedmann spacetimes \cite{ATV2024}. The unique trajectory exiting $SM$ in the opposite overdense direction is the component of the unstable manifold of $SM$ corresponding to $k>0$ Friedmann spacetimes. The primary goal in the present paper is to describe the family $\mathcal{F}$ of solutions of the STV PDE which diverge from one another according to the freedom to assign initial conditions at all orders $n\geq2$ but which agree with a $k<0$ Friedmann spacetime at order $n=1$, when the final SSC gauge freedom associated with time translation is fixed to time since the Big Bang.

\subsection{The Families of Spacetimes $\mathcal{F}_n$ and $\mathcal{F}_{\infty}$}\label{S1.5}

Because the STV ODE of order $n$ are nested and linear in the leading order variables $(z_{2n},w_{2n-2})$, the solutions at order $n$ exist for all time $t>0$ so long as this holds true at order $n-1$ and this statement holds for any choice of initial conditions posed, at every order, at initial time $t=t_0>0$. In particular, we prove that underdense solutions of the STV ODE always decay to $M$ as $t\to\infty$. By compactness, this implies that when time since the Big Bang is assumed, underdense solutions of the STV ODE not only exist, but are bounded for all $t>t_0>0$ at every order $n$, with bounds that depend on the sequence of initial data $(z_{2j}(t_0),w_{2j-2}(t_0))$ for $j=2,\dots,n$ and the choice of $\Delta_0$ at order $n=1$. Based on this, we define $\mathcal{F}_n$ to be the set of solutions of the STV ODE of order $n$ which restricts to a solution which takes $SM$ to $M$ in the phase portrait at order $n=1$ (which we prove then decays to $M$ as $t\to\infty$), with the knowledge that $\mathcal{F}_n$ is a bijective correspondence with any set of initial data assigned at $n\geq2$, together with a choice of $\Delta_0$ to determine the solution on the trajectory which takes $SM$ to $M$ at leading order $n=1$. It follows from the nested property of the STV ODE that solutions in $\mathcal{F}_n$ restrict to solutions in $\mathcal{F}_m$ for every $1\leq m<n<\infty$. More generally, we define $\mathcal{F}_\infty$ to be the set of solutions $\mathcal{F}_n$ defined for all $n\geq1$. Thus a solution in $\mathcal{F}_\infty$ will correspond to an actual smooth solution of the STV ODE on some interval $0<\xi<\xi_0$ if and only if the corresponding expansion in even powers of $\xi$ converges uniformly on compact subsets of that interval. We also define $\mathcal{F}_n'\subset\mathcal{F}_n$ to be the subset of solutions which lie in the unstable manifold of the rest point $SM$ in STV-ODE of order $n$. Note that by definition $\mathcal{F}_1'=\mathcal{F}_1$. Finally, let $\mathcal{F}'\subset \mathcal{F}$ denote the set of solutions of the STV-PDE whose $n^{th}$ order approximation lies in the unstable manifold of rest point $SM$ in the phase portrait of the STV-ODE of order $n$ for every $n\geq1$.

Now it is not difficult to derive sharp bounds on the maximum values $M_n$ of the STV ODE achieved at orders $n\geq1$ at given times $t_n$ sufficient to guarantee an expansion based on coefficients in $\mathcal{F}_\infty$ will converge to a smooth solution of the STV PDE as $n\to\infty$, but the problem of sharply estimating $M_n$ and $t_n$ in terms of the initial data is complicated. In other words, all smooth solutions of the STV PDE in $\mathcal{F}$ will agree with a $k<0$ Friedmann spacetime at order $n=1$ and will project to a unique sequence of time dependent coefficients in $\mathcal{F}_\infty$, but solutions in $\mathcal{F}_\infty$ will not correspond to the expansion of an actual solution of the STV PDE when $n\geq2$ without appropriate estimates satisfied by the initial data for the STV ODE at orders $n\geq2$. Thus we interpret $\mathcal{F}_\infty$ as defining an extension of $\mathcal{F}$ to a set of formal asymptotic expansions in even powers of $\xi$ large enough to allow the imposition of arbitrary initial data for the STV ODE. We call the problem of characterizing the phase portraits of solutions of the STV ODE in $\mathcal{F}_\infty$ the \emph{asymptotics problem} and the problem of finding conditions on initial data sufficient to imply the associated expansions in even powers of $\xi$ will convergence to an exact solution of the STV PDE the \emph{existence problem}. Here we limit ourselves to developing the asymptotic analysis and will address the existence theory in subsequent publications. In summary, we interpret the set of formal expansions in even powers of $\xi$ with coefficients in $\mathcal{F}_\infty$ as generalized solutions of the STV PDE which incorporate all solutions of the STV ODE, this being analogous to, but different from, distributional solutions. Actual solutions of the STV PDE in $\mathcal{F}$ then correspond to the subset of $\mathcal{F}_\infty$ which generate expansions in even powers of $\xi$ which converge uniformly on some interval $0<\xi<\xi_0<1$.

\section{Statement of Results}\label{S2}

Having set out the main elements in words, we now discuss them in mathematical detail.

\subsection{A Self-Similar Expression of the Critical Friedmann Spacetime}\label{S2.1}

To study the stability of the critical ($k=0$) Friedmann spacetime, we look for a change of coordinates which represents it as a stationary solution, so that its stability can be determined by an eigenvalue analysis at a rest point. This has been accomplished for the equation of state $p=\sigma\rho$, with $0\leq\sigma<1$ constant, as follows. To start, recall that the $k=0$ Friedmann metric in comoving coordinates $(t,r)$ takes the form
\begin{align*}
    ds^2 = -dt^2 + R^2(t)(dr^2+r^2d\Omega^2),
\end{align*}
where
\begin{align*}
    d\Omega^2 = d\theta^2 + \sin^2\theta d\theta^2,
\end{align*}
is the standard line element on the unit two-sphere and $R(t)$ is the cosmological scale factor. Assuming time since the Big Bang, so $R(t)=0$ at $t=0$, and assuming $p=\sigma\rho$, then the $k=0$ Friedmann spacetime is given by the exact formulas:
\begin{align}
    R(t) &= t^{\frac{2}{3(1+\sigma)}},\label{R(t)}\\
    H(t) &= \frac{2}{3(1+\sigma)},\label{H(t)}\\
    \rho(t) &= \frac{4}{3\kappa(1+\sigma)^2t^2},\label{rho(t)}
\end{align}
where $H=\dot{R}/R$ is the Hubble function and $\rho$ is the density.

\begin{theorem}\label{T1}
    Let $\eta=\bar{r}/t$ and define the transformation $\Psi:(t,r)\to(\bar{t},\bar{r})$ by
    \begin{align*}
        (\bar{t},\bar{r}) = \big(F(\eta)t,\eta t\big) = \big(F(rt^{\frac{\alpha}{2}-1})t,rt^{\frac{\alpha}{2}}\big),
    \end{align*}
    where
    \begin{align*}
        F(\eta) = \bigg(1+\frac{\alpha(2-\alpha)}{4}\eta^2\bigg)^{\frac{1}{2-\alpha}}, \qquad \alpha = \frac{4}{3(1+\sigma)}.
    \end{align*}
    Then
    \begin{align*}
        \xi = \frac{\bar{r}}{\bar{t}} = \frac{\eta}{F(\eta)},
    \end{align*}
    that is, $\eta$ is an implicit function of $\xi$ alone, $\Psi$ is a bijective regular coordinate transformation for all $\xi<1$ and $\Psi$ transforms the $k=0$ Friedmann metric to the SSC metric form
    \begin{align*}
        ds^2 = -B_\sigma d\bar{t}^2 + \frac{d\bar{r}^2}{A_\sigma} + \bar{r}^2d\Omega^2,
    \end{align*}
    where
    \begin{align*}
        A_\sigma = 1 - \Big(\frac{\alpha\eta}{2}\Big)^2, \qquad
        B_\sigma = \bigg(1-\Big(\frac{\alpha\eta}{2}\Big)^2\bigg)^{-1}\bigg(1+\frac{\alpha(2-\alpha)}{4}\eta^2\bigg)^{\frac{2-2\alpha}{2-\alpha}},
    \end{align*}
    noting that $B_\sigma=1$ for $r=0$. Moreover,
    \begin{align*}
        \kappa\rho_\sigma\bar{r}^2 = \frac{3}{4}\alpha^2\eta^2,
    \end{align*}
    so
    \begin{align*}
        z = \kappa\rho\bar{r}^2, \qquad w = \frac{v}{\xi},
    \end{align*}
    are the density and velocity variables respectively, which depend only on $\xi$.
\end{theorem}

The equations which represent the $k=0$ Friedmann spacetime as a rest point are the Einstein--Euler equations in standard Schwarzschild coordinates using the variables $(\bar{t},\xi)$ and employ $z$ and $w$ as the density and velocity variables respectively. This is accomplished in the next section.

\subsection{The STV PDE}\label{S2.2}

A generic spherically symmetric metric may take, without loss of generality, the following standard Schwarzschild coordinate (SSC) form\footnote{Specifically, under the condition $\frac{\partial C(t,r)}{\partial r}\neq0$, where $C(t,r)d\Omega^2$ is the angular part of the metric, see \cite{STV2017}.}
\begin{align}
    ds^2 = -B(\bar{t},\bar{r})d\bar{t}^2 + \frac{d\bar{r}^2}{A(\bar{t},\bar{r})} + \bar{r}^2d\Omega^2.\label{SSC}
\end{align}
Following \cite{GT2004}, three of the four Einstein field equations determined by $G=\kappa T$ are first order, with the remaining equation second order. The first order equations are given by:\footnote{Metric entries $(A,B)$ are related to $(\hat{A},\hat{B})$ in \cite{GT2004} by $A=1/\hat{B}$, $B=\hat{A}$.}
\begin{align}
    -\bar{r}\frac{A_{\bar{r}}}{A} + \frac{1-A}{A} &= \frac{\kappa B}{A}T^{00}\bar{r}^2,\label{FirstOrder1}\\
    \frac{A_{\bar{t}}}{A} &= \frac{\kappa B}{A}T^{01}\bar{r},\label{FirstOrder2}\\
    \bar{r}\frac{B_{\bar{r}}}{B} - \frac{1-A}{A} &= \frac{\kappa}{A^2}T^{11}\bar{r}^2.\label{FirstOrder3}
\end{align}
The second order equation is given by
\begin{align}
    -\bigg(\frac{1}{A}\bigg)_{\bar{t}\bar{t}} + B_{\bar{r}\bar{r}} - \Phi = \frac{2\kappa B}{A}T^{22}\bar{r}^2,\label{SecondOrder}
\end{align}
where
\begin{align*}
    \Phi = \frac{A_{\bar{t}}B_{\bar{t}}}{2A^2B} - \frac{1}{2A}\bigg(\frac{A_{\bar{t}}}{A}\bigg)^2 - \frac{1}{\bar{r}}B_{\bar{r}} - \frac{BA_{\bar{r}}}{\bar{r}A} + \frac{1}{2}B\bigg(\frac{B_{\bar{r}}}{B}\bigg)^2 - \frac{A_{\bar{r}}B_{\bar{r}}}{2A}.
\end{align*}
The Bianchi identities imply $\nabla\cdot G=0$, so the two conservation laws $\nabla\cdot T=0$ follow identically from $G=\kappa T$. Following \cite{GT2004}, $\nabla\cdot T=0$ is equivalent to:
\begin{align}
    \big(T^{00}_M\big)_{\bar{t}} + \big(\sqrt{AB}T^{01}_M\big)_{\bar{r}} &= -\frac{2}{\bar{r}}\sqrt{AB}T^{01}_M,\label{ConservationLaw1}\\
    \big(T^{01}_M\big)_{\bar{t}} + \big(\sqrt{AB}T^{11}_M\big)_{\bar{r}} &= -\frac{1}{2}\sqrt{AB}\bigg(\frac{4}{\bar{r}}T^{11}_M + \frac{1}{\bar{r}}\Big(\frac{1}{A}-1\Big)\big(T^{00}_M-T^{11}_M\big) + \frac{2\kappa\bar{r}}{A}\big(T^{00}_MT^{11}_M-(T^{01}_M)^2\big) - 4\bar{r}T^{22}\bigg),\label{ConservationLaw2}
\end{align}
where $T_M$ is the Minkowski stress tensor defined by:
\begin{align*}
    T^{00}_M &= BT^{00}, & T^{01}_M &= \sqrt{\frac{B}{A}}T^{01}, & T^{11}_M &= \frac{1}{A}T^{11}, & T^{22}_M &= T^{22}.
\end{align*}
When an equation of state is imposed, the Einstein field equations $G=\kappa T$ for metrics in SSC are weakly equivalent to (\ref{FirstOrder1}), (\ref{FirstOrder3}), (\ref{ConservationLaw1}) and (\ref{ConservationLaw2}). Furthermore, the SSC metric form (\ref{SSC}) is invariant under transformation of the time coordinate, so to restrict this gauge freedom, we say the SSC metric is in \emph{normal gauge} (NG) when geodesic time is imposed at $\bar{r}=0$, that is, we define SSCNG to be the SSC metric (\ref{SSC}) under the NG assumption
\begin{align}
    B(\bar{t},0)=1.\label{B=1}
\end{align}
In SSCNG, the only remaining gauge freedom for the SSC ansatz is time translation freedom $\bar{t}\to\bar{t}+\bar{t}_0$, as this alone preserves (\ref{B=1}).

We now present the STV PDE, a new quasi-self-similar version of the Einstein--Euler equations in SSCNG. The STV PDE are obtained by taking the self-similar variable $\xi=\bar{r}/\bar{t}$ to be the spatial variable in place of $\bar{r}$, under the assumption $|\xi|<1$.

\begin{theorem}[The STV PDE]\label{T2}
    Assume the equation of state $p=\sigma\rho$ with constant $\sigma$. Then for $\xi<1$, the Einstein--Euler equations in SSCNG, expressed in self-similar variables $(\bar{t},\xi)$, are equivalent to the following four equations in unknowns $z(\bar{t},\xi)$, $w(\bar{t},\xi)$, $A(\bar{t},\xi)$ and $D(\bar{t},\xi)$:
    \begin{align}
        \bar{t}z_{\bar{t}} + \xi\big((-1+Dw)z\big)_\xi &= -Dwz,\\
        \bar{t}w_{\bar{t}} + (-1+Dw)\xi w_\xi &- w + Dw^2 - \frac{(1+\sigma^2)\sigma^2}{\xi z}\bigg(D\frac{(1-v^2)v^2}{(1+\sigma^2v^2)^2}z\bigg)_\xi+ \frac{\sigma^2\xi}{z}\bigg(D\frac{1-v^2}{1+\sigma^2v^2}\frac{z}{\xi^2}\bigg)_\xi\notag\\
        &= -\frac{1-v^2}{1+\sigma^2v^2}\frac{D}{2\xi^2A}\bigg((1-\sigma^2)(1-A)+2\sigma^2\frac{1-v^2}{1+\sigma^2v^2}z\bigg),\\
        \xi A_\xi &= -z + (1-A),\\
        \xi D_\xi &= \frac{D}{2A}\bigg(2(1-A)-(1-\sigma^2)\frac{1-v^2}{1+\sigma^2 v^2}z\bigg),
    \end{align}
    where:
    \begin{align*}
        z &= \frac{\kappa\rho\bar{r}^2}{1-v^2}, & w &= \frac{1+\sigma^2}{1+\sigma^2v^2}\frac{v}{\xi}, & D &= \sqrt{AB}.
    \end{align*}
\end{theorem}

Our concern in this paper is with the case $p=\sigma=0$, applicable to late time Big Bang Cosmology.\footnote{In the Standard Model of Cosmology, the pressure drops precipitously to zero at about 10,000 years after the Big Bang, an order of magnitude before the uncoupling of matter and radiation \cite{P1999}.} In this case, the STV PDE reduce to the following simpler system, which is the subject of this paper:
\begin{align}
    \bar{t}z_{\bar{t}} + \xi\big((-1+Dw)z\big)_{\xi} &= -Dwz,\label{p=0z}\\
    \bar{t}w_{\bar{t}} + \xi(-1+Dw)w_{\xi} &= w-D\bigg(w^2+\frac{1}{2\xi^2}(1-\xi^2w^2)\frac{1-A}{A}\bigg),\label{p=0w}\\
    \xi A_{\xi} &= -z + (1-A),\label{p=0A}\\
    \xi D_{\xi} &= \frac{D}{2A}\big(2(1-A)-(1-v^2)z\big).\label{p=0D}
\end{align}

The authors' original motivation to formulate a self-similar version of the Einstein--Euler equations was the discovery that, when $p=\sigma\rho$ and the usual gauge of geodesic time at $\bar{r}=0$ with the Big Bang at $\bar{t}=0$ is employed, the $k=0$ Friedmann spacetime in SSC has the property that all of the variables $z$, $w$, $A$ and $D$ are functions of $\xi=\bar{r}/\bar{t}$ alone, and hence represent a time independent solution, or \emph{rest point}, of the STV PDE, suggesting to the authors that such a PDE would be useful in studying the stability properties of the Standard Model of Cosmology.

\subsection{The STV ODE}\label{S2.3}

The STV ODE are derived by expanding smooth solutions of the STV PDE in powers of $\xi$ with time dependent coefficients, then collecting like powers of $\xi$. The assumption of smoothness at the center implies that the non-zero coefficients in the expansion occur only for even powers $\xi^{2n}$ and this significantly reduces the solution space of the Einstein field equations by disentangling solutions smooth at the center from the larger generic solution space. Fortuitously, the resulting equations in the fluid variables $z$ and $w$ uncouple from the equations for the metric components $A$ and $D$. The expansions take the form:
\begin{align}
    z(\bar{t},\xi) &= z_2(\bar{t})\xi^2 + z_4(\bar{t})\xi^4 + \dotsc + z_{2n}(\bar{t})\xi^{2n} + \dotsc,\label{zExpansion}\\
    w(\bar{t},\xi) &= w_0(\bar{t}) + w_2(\bar{t})\xi^2 + \dotsc + w_{2n-2}(\bar{t})\xi^{2n-2} + \dotsc,\label{wExpansion}\\
    A(\bar{t},\xi) &= 1 + A_2(\bar{t})\xi^2 + A_4(\bar{t})\xi^4 + \dotsc + A_{2n}(\bar{t})\xi^{2n} + \dotsc,\label{AExpansion}\\
    D(\bar{t},\xi) &= 1 + D_2(\bar{t})\xi^2 + D_4(\bar{t})\xi^4 + \dotsc + D_{2n}(\bar{t})\xi^{2n} + \dotsc.\label{DExpansion}
\end{align}
Here we restrict to the case $\sigma=0$. Substituting (\ref{zExpansion})--(\ref{DExpansion}) into equations (\ref{p=0z})--(\ref{p=0D}), we find that the resulting ODE close in $(z_{2k},w_{2k-2})$ for $1\leq k\leq n$ at every order $n\geq1$ when $\sigma=0$. We record the result in the following theorem \cite{ATV2024,STV2017}.

\begin{theorem}[The STV ODE]\label{T3}
    Assume a smooth solution $\big(z(\bar{t},\xi),w(\bar{t},\xi),A(\bar{t},\xi),D(\bar{t},\xi)\big)$ of the STV PDE is expanded in even powers of $\xi$. Then $A_{2k}$ can be re-expressed in terms of $z_{2k}$, and $D_{2k}$ can be re-expressed in terms of $z_{2k}$ and $w_0,\dots,w_{2k-2}$, to form, at each order $n\geq1$, a $2n\times2n$ system of ODE
    \begin{align}
        \bar{t}\dot{\boldsymbol{U}} = \boldsymbol{F}_n(\boldsymbol{U})\label{nxnSystem1}
    \end{align}
    in unknowns
    \begin{align*}
        \boldsymbol{U} = (\boldsymbol{v}_1,\dots,\boldsymbol{v}_n)^T,
    \end{align*}
    where
    \begin{align*}
        \boldsymbol{v}_{k}=(z_{2k},w_{2k-2})^T.
    \end{align*}
    Moreover, system (\ref{nxnSystem1}) takes the component form
    \begin{align}
        \bar{t}\frac{d}{d\bar{t}}\left(\begin{array}{c}
        \boldsymbol{v}_1\\
        \boldsymbol{v}_2\\
        \vdots\\
        \boldsymbol{v}_n\end{array}\right) = \left(\begin{array}{c}
        P_1\boldsymbol{v}_1+\boldsymbol{q}_1\\
        P_2\boldsymbol{v}_2+\boldsymbol{q}_2\\
        \vdots\\
        P_n\boldsymbol{v}_n+\boldsymbol{q}_n
        \end{array}\right),\label{nxnSystem2}
    \end{align}
    where
    \begin{align}
        P_k = P_k(\boldsymbol{v}_1) = \left(\begin{array}{cc}
        (2k+1)(1-w_0)-1 & -(2k+1)z_2\\
        -\frac{1}{2(2k+1)} & 2k(1-w_0)-1
        \end{array}\right)\label{PkMatrix}
    \end{align}
    depends only on $\boldsymbol{v}_1=(z_2,w_0)^T$ and:
    \begin{align*}
        \boldsymbol{q}_1 &= \boldsymbol{q}_1(\boldsymbol{v}_1),\\
        \boldsymbol{q}_k &= \boldsymbol{q}_k(\boldsymbol{v}_1,\dots,\boldsymbol{v}_{k-1}).
    \end{align*}
\end{theorem}

It follows that the STV ODE are nested in the sense that each STV ODE of order $n\geq2$ contains as a sub-system the STV ODE of order $k$ for all $1\leq k\leq n-1$. Thus the self-similar formulation decouples solutions at every order, which is to say that one can solve for solutions up to order $n-1$ and use the STV ODE at order $n$ to solve for $(z_{2n}(\bar{t}),w_{2n-2}(\bar{t}))$ from arbitrary initial conditions $(z_{2n}(0),w_{2n-2}(0))$. Assuming lower order solutions $k\leq n-1$ are fixed, the STV ODE of order $n$ turn out to be linear in the highest order terms $(z_{2n},w_{2n-2})$. For approximations up to order $n=2$ we can use the approximation $z=\kappa\rho\bar{r}^2$ because this incurs errors of $O(\xi^6)$ given that $v^2=O(\xi^2)$.

Each STV ODE of order $n\geq1$ admit the rest points $SM$ and $M$. The coordinates of the rest point $SM$ are obtained by expanding the self-similar formulation of the $k=0$ Friedmann spacetime in even powers of $\xi$ about the center. This is because the $k=0$ Friedmann spacetime is a time independent solution of the STV PDE and it follows that the resulting expansion gives the coordinates of the rest point $SM$ at every order. The rest point $M$, which represents Minkowski spacetime (and the limit $\rho\to0$, $v\to\xi$), has zeros in all entries except for a $1$ in the (second) $w_0$-entry, that is, $M=(0,1,0,\dots,0)$. It is easy to verify $M$ is a rest point by induction using (\ref{nxnSystem2}). We record the rest points $SM$ and $M$, together with their eigenpairs, as follows.

\begin{corollary}\label{C1}
    Each STV ODE of order $n\geq1$ admit the rest points:
    \begin{align}
        M = (0,1,0,\dots,0)\in\mathbb{R}^{2n}, \qquad SM = \bigg(\frac{4}{3},\frac{2}{3},\frac{40}{27},\frac{2}{9},\dots\bigg)\in\mathbb{R}^{2n}.
    \end{align}
    The rest point $M$ is a degenerate stable rest point with one eigenvalue and one eigenvector given by:
    \begin{align}
        \lambda_M = -1, \qquad \boldsymbol{R}_M = (0,1,0,1,\dots)^T.\label{EigenvaluesM}
    \end{align}
    The eigenvalues of $SM$, computable directly from (\ref{PkMatrix}), are given by:
    \begin{align}
        \lambda_{An} = \frac{2n}{3}, \qquad \lambda_{Bn} = \frac{1}{3}(2n-5).\label{EigenvaluesSM}
    \end{align}
    The corresponding eigenvectors up to order $n=2$ are given by:
    \begin{align}
        \lambda_{A1} &= \frac{2}{3}, & \boldsymbol{R}_1 := \boldsymbol{R}_{A1} &= \bigg(-9,\frac{3}{2},-\frac{10}{3},-1\bigg)^T;\\
        \lambda_{B1} &= -1, & \boldsymbol{R}_{B1} &= \bigg(4,1,\frac{80}{9},1\bigg)^T;\\
        \lambda_{A2} &= \frac{4}{3}, & \boldsymbol{R}_3 := \boldsymbol{R}_{A2} &= (0,0,-10,1)^T;\\
        \lambda_{B2} &= -\frac{1}{3}, & \boldsymbol{R}_{B2} &= \bigg(0,0,\frac{20}{3},1\bigg)^T.
    \end{align}
\end{corollary}

\subsection{The STV ODE Phase Portrait of Order $n=1$}\label{S2.4}

\begin{figure}[hbt!]
    \centering
    \caption{The phase portrait for the $2\times2$ system.}
    \includegraphics[width=0.8\textwidth]{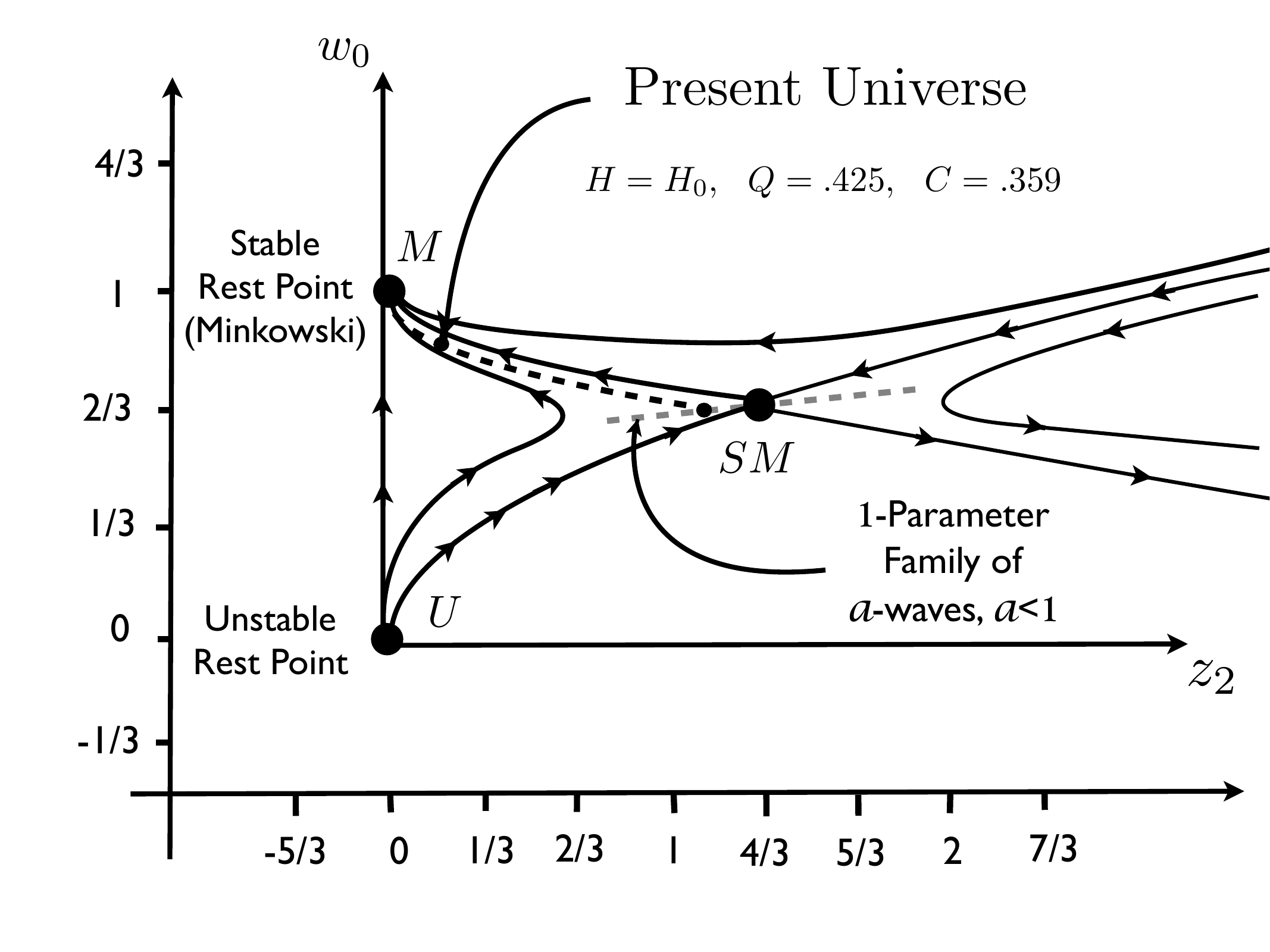}\label{Figure1}
\end{figure}

The STV ODE of order $n=1$ is the $2\times2$ system:\footnote{Note that in \cite{STV2017}, the fixed point $SM$ of the $n=1$ phase portrait was centered on $(0,0)$.}
\begin{align}
    \bar{t}\dot{z}_2 &= 2z_2 - 3z_2w_0,\label{z2ODE2x2}\\
    \bar{t}\dot{w}_0 &= -\frac{1}{6}z_2 + w_0 - w_0^2.\label{w0ODE2x2}
\end{align}
Using
\begin{align*}
    \frac{d}{d\tau} = \bar{t}\frac{d}{d\bar{t}},
\end{align*}
equations (\ref{z2ODE2x2}) and (\ref{w0ODE2x2}) convert to an autonomous $2\times2$ system of ODE in $\tau=\ln\bar{t}$. From Corollary \ref{C1} or by direct calculation, we see that the system admits three rest points: The source $U=(0,0)$, the unstable saddle rest point $SM=(\frac{4}{3},\frac{2}{3})$ and the degenerate stable rest point $M=(0,1)$. The rest point $U$ plays no role once time since the Big Bang is imposed, the rest point $M$ describes the asymptotics of solutions tending to Minkowski spacetime as $\bar{t}\to\infty$ and the coordinates of rest point $SM$ are precisely the first two terms in the self-similar expansion of the $k=0$ Friedmann spacetime, viewed here as the Standard Model due to the central role it has played in the history of Cosmology. The first two components of $\boldsymbol{R}_M$ give the single eigenvector at $M$, with the first two components of $\boldsymbol{R}_{A1}$ and $\boldsymbol{R}_{B1}$ giving the eigenvectors of $SM$. The phase portrait for system (\ref{z2ODE2x2})--(\ref{w0ODE2x2}) is diagrammed in Figure \ref{Figure1}\footnote{Figure \ref{Figure1} is a modification of Figure 1 from \cite{STV2017}}. The two components of the unstable manifold of $SM$ correspond to the two trajectories associated with the positive eigenvalue $\lambda_{A1}=\frac{2}{3}$, the underdense component being the trajectory which connects $SM$ to $M$ and the overdense component leaving $SM$ in the opposite direction, see Figure \ref{Figure1}. The trajectory connecting $U$ to $SM$ is the underdense trajectory in the stable manifold of $SM$ corresponding to the negative eigenvalue $\lambda_{B1}=-1$ and the overdense stable trajectory emerges opposite to this at $SM$.

\subsection{The Trajectory Connecting $SM$ to $M$}\label{S2.5}

Using classical formulas for $k\neq0$ Friedmann spacetimes, we verify that the underdense trajectory connecting $SM$ to $M$ at order $n=1$ corresponds to $k<0$ Friedmann spacetimes and the overdense trajectory in the unstable manifold of $SM$ corresponds to $k>0$ Friedmann spacetimes. This is proven by converting implicit formulas for Friedmann spacetimes in comoving coordinates over to SSCNG, in particular, converting the resulting formulas into expressions for $z$ and $w$ in terms of $(\bar{t},\xi)$, expanding these in even powers of $\xi$ and then extracting the leading orders. This extends to a procedure for obtaining exact formulas for the trajectories in the unstable manifold of $SM$ at every order $n\geq1$. We summarize the procedure through the following theorems, starting with known formulas for Friedmann spacetimes in comoving coordinates \cite{GH2007}.

\begin{theorem}\label{T4}
    Recall that the one parameter family of Friedmann metrics is given in comoving coordinates by
    \begin{align*}
        ds^2 = -dt^2 + \frac{R^2(t)}{1-kr^2}dr^2 + R^2(t)r^2d\Omega^2,
    \end{align*}
    where $R(t)$ is the cosmological scale factor and $k$ the curvature parameter. Then the coordinate transformation $(\bar{t},\bar{r})=\Psi(t,r)$ given by
    \begin{align*}
        (\bar{t},\bar{r}) = \big(F(h(t)g(r)),R(t)r\big),
    \end{align*}
    where:
    \begin{align*}
        h(t) &= e^{\lambda\int_0^t\frac{d\tau}{\dot{R}(\tau)R(\tau)}},\\
        g(r) &= \begin{cases}
        (1-kr^2)^{-\frac{\lambda}{2k}}, & k\neq0,\\
        e^{\frac{\lambda}{2}r^2}, & k=0,
        \end{cases}\\
        F(y) &= h^{-1}(y),
    \end{align*}
    converts the Friedmann spacetime from comoving coordinates to SSC for each $\lambda>0$. Moreover, taking $\lambda=\frac{1}{2}$, the resulting SSCNG metric is given by:
    \begin{align}
        A &= 1 - kr^2 - H^2\bar{r}^2,\label{SSCNG:A}\\
        B &= \frac{1}{F'(\Phi)^2}\hat{B} = \frac{1}{(F'(\Phi)\Phi_t)^2}\frac{1-kr^2}{1-kr^2-H^2\bar{r}^2},\label{SSCNG:B}
    \end{align}
    where $\Phi(t,r) = f(t)g(r)$ and
    \begin{align}
        \sqrt{AB} &= \frac{\sqrt{1-kr^2}}{\frac{\partial\bar{t}}{\partial t}(t,r)},\label{sqrtABbest}\\
        v &= \frac{\dot{R}r}{\sqrt{1-kr^2}},\label{SSCNG:v}
    \end{align}
    with $v$ being the SSCNG coordinate velocity.\footnote{Note that (\ref{SSCNG:A}) and (\ref{SSCNG:B}) agree with equation (2.19) of \cite{ST2004}.}
\end{theorem}

The next theorem uses implicit formulas for Friedmann spacetimes in comoving coordinates (see \cite{GH2007}) to derive an exact formula for the orbit connecting $SM$ to $M$ in the $n=1$ phase portrait. This establishes that the trajectory corresponds to the leading order part of the $k=-1$ Friedmann spacetime, with the free parameter $\Delta_0$ entering in the log-time translation $\tau-\Delta_0$ associated with trajectories of the autonomous STV ODE.

\begin{theorem}\label{T5}
    An exact formula for the $p=0$, $k=-1$ Friedmann spacetime in comoving coordinates is given in \cite{GH2007} by
    \begin{align*}
        R = \frac{\Delta_0}{2}(\cosh2\theta-1),
    \end{align*}
    where $\theta=\theta\big(\frac{t}{\Delta_0}\big)$ is the function determined by inverting the relation
    \begin{align*}
        \frac{t}{\Delta_0} = \frac{1}{2}(\sinh2\theta-2\theta).
    \end{align*}
    Transforming to SSCNG by Theorem \ref{T4}, expanding $z$ and $w$ in powers of $\xi$ using (\ref{SSCNG:A})--(\ref{SSCNG:v}) and extracting the leading order terms leads to the following exact formula for the trajectory $\big(z_2^F(\bar{t}),w_0^F(\bar{t})\big)$ connecting $SM$ to $M$ in phase portrait of the STV ODE of order $n=1$:
    \begin{align}
        \big(z_2^F(\bar{t}),w_0^F(\bar{t})\big) = \big(\tilde{z}_2(\theta),\tilde{w}_0(\theta)\big),\label{ExactOrbit}
    \end{align}
    where
    \begin{align}
        \big(\tilde{z}_2(\theta),\tilde{w}_0(\theta)\big) = \bigg(\frac{6(\sinh2\theta-2\theta)^2}{(\cosh2\theta-1)^3},\frac{(\sinh2\theta-2\theta)\sinh2\theta}{(\cosh2\theta-1)^2}\bigg),
    \end{align}
    $\theta=\theta\big(\frac{t}{\Delta_0}\big)$ defines $\theta$ as a function of $t$ and $t$ is given implicitly as a function of of $\bar{t}$ through the relation
    \begin{align}
        \cosh\bar{\theta} = \sqrt[4]{1+\frac{\bar{t}^2\xi^2}{\Delta_0^2\sinh^4\theta}}\cosh\theta,
    \end{align}
    where $\bar{\theta}=\theta\big(\frac{\bar{t}}{\Delta_0}\big)$. Furthermore, for $n=2$ we have
    \begin{align}
        \big(\tilde{z}_4(\theta),\tilde{w}_2(\theta)\big) = \bigg(\frac{30(\sinh2\theta-2\theta)^4\cosh^2\theta}{(\cosh2\theta-1)^6},\frac{3(\sinh2\theta-2\theta)^3\cosh\theta}{2(\cosh2\theta-1)^4\sinh\theta}\bigg).
    \end{align}
\end{theorem}

The following result asserts that self-similar coordinates are valid for $k<0$ Friedmann spacetimes out to approximately the Hubble radius $\xi=1$.

\begin{theorem}\label{T6}
    The transformation $(t,r)\to(\bar{t},\xi)$, defined for the $p=0$, $k=-1$ Friedmann spacetime, is a regular bijective transformation of the SSC $(\bar{t},\bar{r})$ to SSC self-similar $(\bar{t},\xi)$ coordinates for all $\xi\leq\xi_0\approx0.816$. Moreover, this region of validity extends to $\xi<1$ as $\bar{t}\to\infty$.
\end{theorem}

\subsection{Time Since The Big Bang}\label{S2.6}

\begin{figure}[hbt!]
    \centering
    \caption{The space $\mathcal{F}$ of solutions which decay to $M$.}
    \includegraphics[width=0.9\textwidth]{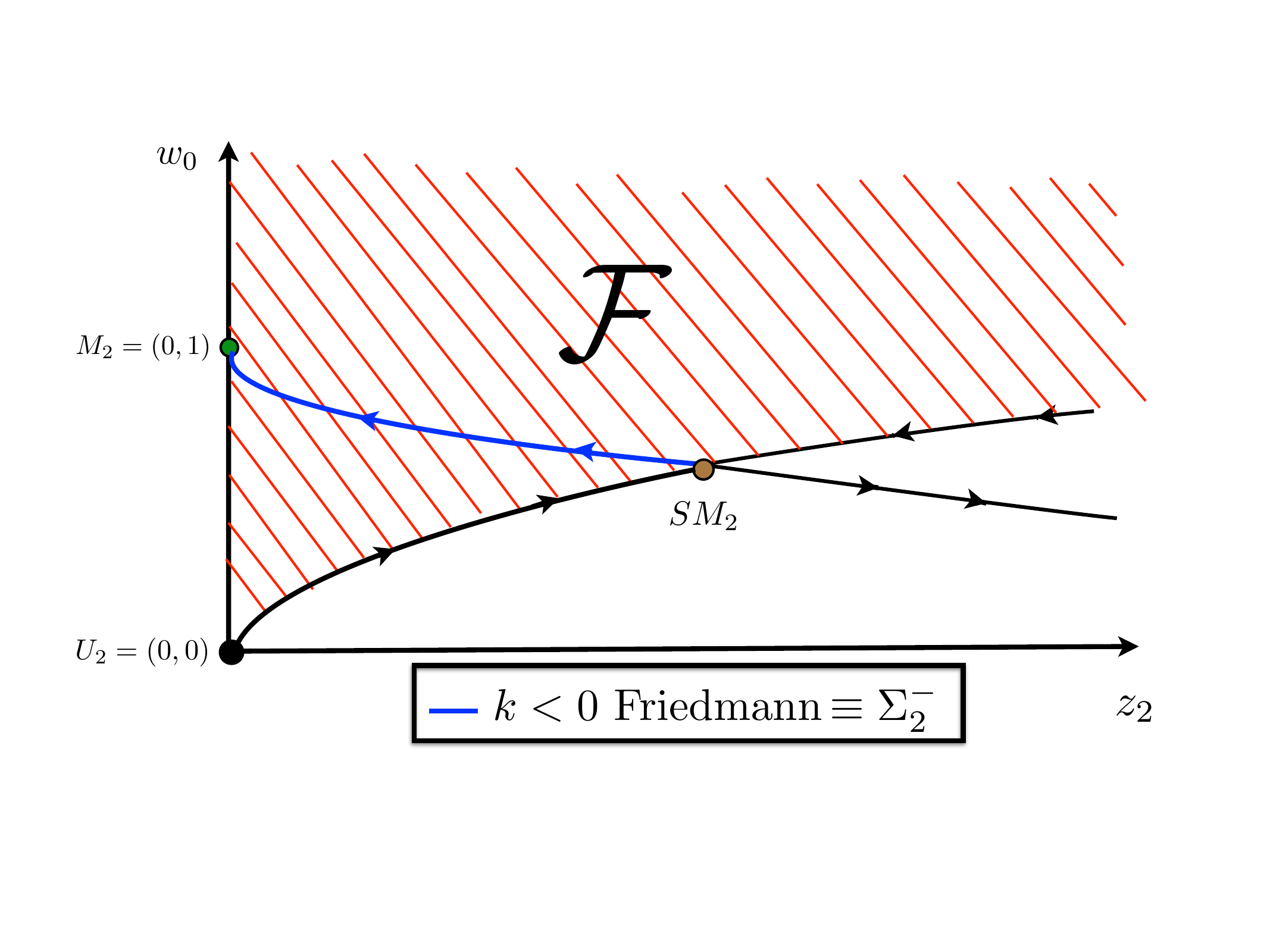}\label{Figure2}
\end{figure}

Standard Schwarzschild coordinates with the normal gauge still leaves one gauge freedom yet to be set, namely, the freedom to impose time translation $\bar{t}\to\bar{t}-\bar{t}_0$. The time translation freedom of the SSC system leaves open an unresolved redundancy in solutions of the STV ODE in the sense that time translation maps each trajectory of the STV ODE of order $n=1$ to a different trajectory even though they represent the same physical solution. We show for each trajectory of the system (\ref{z2ODE2x2})--(\ref{w0ODE2x2}) that there exists a unique time translation $\bar{t}\to\bar{t}-\bar{t}_0$, which we call \emph{time since the Big Bang}, and which converts that trajectory either to $SM$ or to one of the two trajectories in the unstable manifold of $SM$. In particular, referring to the phase portrait depicted in Figure \ref{Figure1} and making the gauge transformation to time since the Big Bang, the trajectories in the stable manifold of $SM$, that is, the one taking $U$ to $SM$ and the trajectory opposite it at $SM$, go over to $SM$, whereas all the trajectories above these, the trajectories in the domain of attraction of $M$, go over to the underdense portion of the unstable manifold of $SM$ corresponding to $k<0$ Friedmann spacetimes. Trajectories below the stable manifold of $SM$ go over to the overdense portion of the unstable manifold of $SM$, corresponding to $k>0$ Friedmann spacetimes. This is depicted in Figure \ref{Figure2}. From this it follows that imposing the solution dependent time since the Big Bang has the effect of eliminating the negative eigenvalue $\lambda_{B1}=-1$ together with the rest point $U$, and we can, without loss of generality, restrict our analysis to the space of solutions which agree with $SM$ or a trajectory in its unstable manifold, in the phase portrait of the solution at $n=1$. Since these trajectories agree with the Friedmann spacetimes, we conclude that, under appropriate change of time gauge, all smooth solutions of the STV PDE agree with a Friedmann spacetime at leading order in the STV ODE. Our purpose here is to study the space $\mathcal{F}$ of solutions which lie on the trajectory which takes $SM$ to $M$ just at leading order, that is, the set of all solutions which agree with a $k<0$ Friedmann spacetime at order $n=1$ of the STV ODE. We do not consider the $k>0$ Friedmann spacetimes but observe that these exit the first quadrant of our coordinate system at $w_0=0$, the time of maximal expansion.\footnote{The $k>0$ case is complicated by the fact that the time of maximal expansion, when $v=0$, need not be the time when any single $v_n$ is zero at order $n\geq0$.}

The rest point $M$ describes the time asymptotic decay of solutions in $\mathcal{F}$ to Minkowski spacetime as $\bar{t}\to\infty$. A calculation shows that $M$ is a degenerate stable rest point with repeated eigenvalue $\lambda_M=-1$ and single eigenvector $\boldsymbol{R}_M=(0,1)^T$. Thus solutions in $\mathcal{F}$ decay to $M$ time asymptotically along the $w_0$-axis as $\bar{t}\to\infty$. Moreover, as is standard for degenerate stable rest points with the character of $M$, $z_2(\bar{t})$ and $w_0(\bar{t})$ decay to $M$ at leading order like $O(\bar{t}^{-1})$ and $O(\bar{t}^{-1}\ln\bar{t})$ respectively. Thus, assuming solutions in $\mathcal{F}$ agree with a Friedmann spacetime at leading order, but diverge at higher orders with errors estimated by Taylor's theorem, we can use the $n=1$ phase portrait of the STV ODE alone to conclude that, to leading order as $\bar{t}\to\infty$, every solution in $\mathcal{F}$ decays to $w=v/\xi=1$ and $z=0$ at the same rate for fixed $\xi$ and decays to a Friedmann spacetime faster than to Minkowski spacetime for fixed $\bar{r}>0$ (since $\xi\to0$). Corollary \ref{C1} establishes that $M$ is a degenerate stable rest point in the phase portrait of the STV ODE at every order $n\geq1$, exhibiting the same negative eigenvalue $\lambda_{M}=-1$ with a single eigenvector $\boldsymbol{R}_M$ at each order. The degenerate structure of the rest point $M$ at all orders implies that the estimate for the discrepancy between a solution in $\mathcal{F}$ and the Friedmann spacetime it agrees with at leading order, is estimated by the discrepancy at second order as $\bar{t}\to\infty$. We conclude that one would see perfect alignment between solutions in $\mathcal{F}$ and $k<0$ Friedmann spacetimes at order $n=1$, and the error between them tends to zero by a factor $O(\bar{t}^{-1})$ faster as $\bar{t}\to\infty$ (for each fixed $\bar{r}>0$) than what one would see without taking account of the decay of solutions to $M$ at higher orders. The result, which uses standard rates of decay at degenerate stable rest points with the character of $M$, is recorded in the following theorem.

\begin{theorem}\label{T7}
    Imposing time since the Big Bang, each solution in $\mathcal{F}$ agrees exactly with a single $k<0$ Friedmann spacetime at leading order in the STV ODE. Moreover, the discrepancy between a general solution in $\mathcal{F}$ and the $k<0$ Friedmann spacetime it agrees with at leading order is estimated at fixed $\bar{r}<\bar{t}$ by:
    \begin{align}
        |\rho(\bar{t},\bar{r})-\rho_F(\bar{t},\bar{r})| &\leq C(\bar{r}) \bar{t}^{-5},\label{rhoDecay1}\\
        |v(\bar{t},\bar{r})-v_F(\bar{t},\bar{r})| &\leq C(\bar{r}) \bar{t}^{-4}\ln\bar{t}\label{vDecay1},
    \end{align}
    where $C(\bar{r})$ depends on $\bar{r}<\bar{t}$ and the particular solution. The decay to $M$ implies the rate of decay to this Friedmann spacetime is faster than the rate of decay to Minkowski spacetime, which is estimated at fixed $\bar{r}$ by:
    \begin{align}
        |\rho(\bar{t},\bar{r})| &\leq C(\bar{r})\bar{t}^{-3},\label{rhoDecay2}\\
        |v(\bar{t},\bar{r})-\xi| &\leq C(\bar{r})\bar{t}^{-2}\ln\bar{t}.\label{vDecay2}
    \end{align}
    Here (\ref{rhoDecay1})--(\ref{vDecay2}) hold not just as $\bar{t}\to\infty$, but also globally for any solution in $\mathcal{F}$ (and any $\bar{r}<\bar{t}$) by letting $C(\bar{r})$ depend on the initial data and the global bound on the particular solution, a bound which exists due to the decay to $M$ as $\bar{t}\to\infty$.
\end{theorem}

Note that Theorem \ref{T7} gives a faster rate of decay for a $k<0$ Friedmann spacetime over the $O(\bar{t}^{-2})$ decay rate known for the $k=0$ Friedmann spacetime, see (\ref{rho(t)}). From this we conclude faster decay to Friedmann spacetimes than to Minkowski spacetime and faster decay in the density than in the velocity, which tends to $\xi=\bar{r}/\bar{t}$ on approach to the rest point $M$.

An exact formula for the trajectory taking $SM$ to $M$ at order $n=1$ is given by (\ref{ExactOrbit}), and since time since the Big Bang places all solutions smooth at the center precisely on this trajectory, it follows that (\ref{ExactOrbit}) is a canonical leading order evolution shared by all underdense solutions in $\mathcal{F}$, including $k<0$ Friedmann spacetimes. The rest point $SM$ persists to every order because the $k=0$ Friedmann spacetime is self-similar at every order of the STV ODE. Somewhat surprisingly, the crucial behavior of solutions in $\mathcal{F}$ emerges at order $n=2$.

\subsection{The STV ODE Phase Portrait of Order $n=2$}\label{S2.7}

The corrections to $k<0$ Friedmann spacetimes accounted for by solutions in $\mathcal{F}$ at order $n=2$ are most important as this is the order in which a negative eigenvalue emerges at $SM$ and also the leading order in which divergence from Friedmann spacetimes is observed. Furthermore, the order $n=2$ is important because it determines $w_2(\bar{t})\xi^2$, which provides the third order correction to redshift vs luminosity, the correction at the order of the anomalous acceleration of the galaxies which are purportedly accounted for by dark energy in the standard $\Lambda$CDM model of Cosmology \cite{STV2017}. This is discussed further in Section \ref{S2.9}.

\begin{figure}[hbt!]
    \caption{The phase portrait for the $4\times4$ system.}
    \includegraphics[width=\textwidth]{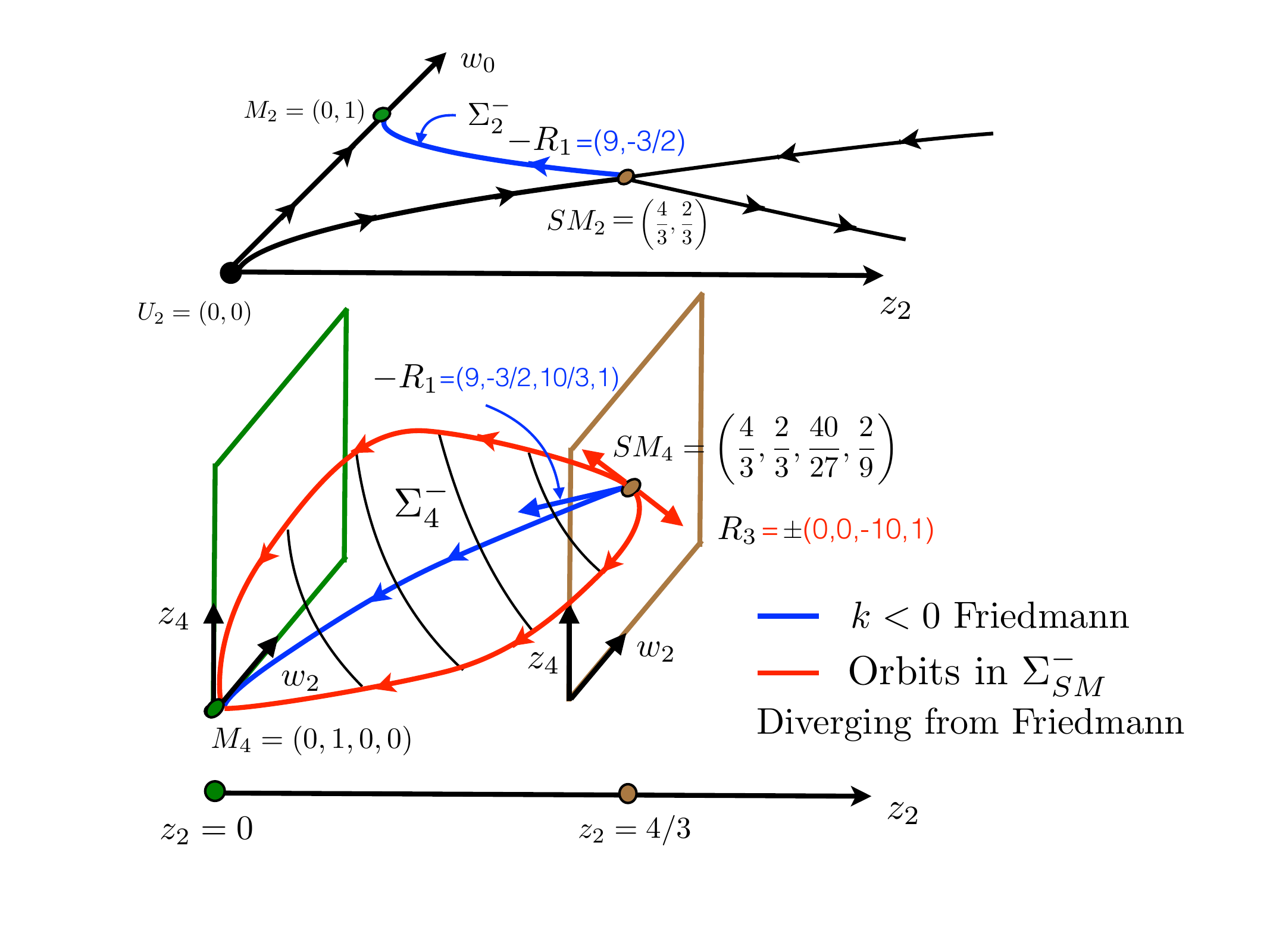}\label{Figure3}
\end{figure}

The STV ODE of order $n=2$ is the $4\times4$ system:\footnote{Note that this corrects an error in \cite{STV2017} in the terms involving $z_4$, a mistake at fourth order in $\xi$ which did not affect the conclusions.}
\begin{align}
    \bar{t}\dot{z}_2 &= 2z_2 - 3z_2w_0,\label{z2ODE4x4}\\
    \bar{t}\dot{w}_0 &= -\frac{1}{6}z_2 + w_0 - w_0^2,\label{w0ODE4x4}\\
    \bar{t}\dot{z}_4 &= \frac{5}{12}z_2^2w_0 + 4z_4 - 5w_0z_4 - 5z_2w_2,\label{z4ODE4x4}\\
    \bar{t}\dot{w}_2 &= -\frac{1}{24}z_2^2 + \frac{1}{4}z_2w_0^2 - \frac{1}{10}z_4 + 3w_2 - 4w_0w_2.\label{w2ODE4x4}
\end{align}
Note first that the STV ODE of order $n=1$ appears as the subsystem (\ref{z2ODE4x4})--(\ref{w0ODE4x4}). Viewed as a $4\times4$ autonomous system, equations (\ref{z2ODE4x4})--(\ref{w2ODE4x4}) admit the three rest points: $U=(0,0,0,0)$, $M=(0,1,0,0)$ and $SM=(\frac{4}{3},\frac{2}{3},\frac{40}{27},\frac{2}{9})$. Imposing time since the Big Bang restricts the solution space to solutions in the unstable manifold of $SM$ at order $n=1$ and this eliminates $U$ from the solution space. The eigenvalues and eigenvectors of rest points $M$ and $SM$ in the STV ODE of order $n=4$ are given by Corollary \ref{C1}. The phase portrait for solutions of (\ref{z2ODE4x4})--(\ref{w2ODE4x4}) is depicted in Figure \ref{Figure3}. Note that $SM$ and $M$ are referred to as $SM_2$ (and $SM_4$) and $M_2$ (and $M_4$) in Figure \ref{Figure3} respectively to indicate that the fixed points are those for the $2\times2$ (and $4\times4$) system.

The time since the Big Bang gauge is assumed in Figure \ref{Figure3}, so all elements of $\mathcal{F}$ agree with a $k<0$ Friedmann spacetime at leading order, that is, they lie on the unstable trajectory taking $SM$ to $M$ in the leading order phase portrait associated with (\ref{z2ODE4x4})--(\ref{w0ODE4x4}). This is denoted by $\Sigma_{SM}^-$ in Figure \ref{Figure3}, with $\Sigma_2^-$ and $\Sigma_4^-$ specifying the unstable manifold for the $2\times2$ and $4\times4$ systems respectively. The phase portraits of Figures \ref{Figure1} and \ref{Figure3} are consistent because the first two components of $-\boldsymbol{R}_1$ give the direction of the trajectory which connects $SM_2$ to $M_2$ at level $n=1$ and the second two components represent the higher order corrections. The projections of the $k<0$ Friedmann spacetimes onto solutions of the STV ODE of orders $n=1$ and $n=2$ are represented by the blue curves in Figure \ref{Figure3}. Note that the presence of a second positive eigenvalue $\lambda_{A2}=\frac{4}{3}$ implies the unstable manifold of $SM$ intersects $\mathcal{F}$ in a two-dimensional space of trajectories emanating from $SM$. We prove that only the eigensolution of $\lambda_{A1}$ corresponds to the Friedmann spacetime at order $n=2$. This immediately implies that there exist solutions in the unstable manifold of $SM$ which agree with a $k<0$ Friedmann spacetime at order $n=1$ but diverge from a Friedmann spacetime at intermediate times. Since all solutions in $\mathcal{F}$ decay to $M$ as $\bar{t}\to\infty$, this implies that solutions in the unstable manifold of $SM$ agree with $k<0$ Friedmann spacetimes in the limits $\bar{t}\to0$, $\bar{t}\to\infty$ (for each fixed $\bar{r}>0$) and at leading order $n=1$, but which diverge, and hence introduce accelerations away from Friedmann spacetimes, at intermediate times. By the Hartman--Grobman theorem, nonlinear solutions correspond to linearized solutions in a neighborhood of a rest point, so solutions in the unstable manifold of $SM$ are determined by their limiting eigendirection $\boldsymbol{R}=a\boldsymbol{R}_1+b\boldsymbol{R}_3$ at $SM$, and hence we conclude that the magnitude of the acceleration away from Friedmann spacetimes is measured by $b/a$, which can be arbitrarily large. More precisely, all solutions in the unstable manifold of $SM$ at order $n=2$ leave $SM$ tangent to:
\begin{align}
    \boldsymbol{U}(\tau) = ae^{\lambda_{A1}(\tau-\tau_0)}\boldsymbol{R}_{A1} + be^{\lambda_{A2}(\tau-\tau_0)}\boldsymbol{R}_{A2}.
\end{align}
Note that the smallest positive eigenvalue to emerge at any order at $SM$ is $\lambda_{B3}=\frac{1}{3}$, and since all trajectories in $\mathcal{F}$ have a non-zero component of $\boldsymbol{R}_{A1}$ by definition, we can further conclude that all trajectories in the unstable manifold of $SM$ are tangent to $\boldsymbol{R}_{A1}$ in backward time at $SM$ in the phase portrait of the STV ODE of order $n=2$ but come in tangent to $\boldsymbol{R}_{B3}$ at $SM$ in the portraits of the STV ODE at all higher orders. We show that the $k<0$ Friedmann spacetimes have no components in direction $\boldsymbol{R}_{B3}$ and that by the nested structure of the STV ODE, $\boldsymbol{R}_{Bn}$ has non-zero components in only leading order entries.

The existence of a unique negative eigenvalue $\lambda_{B2}=-\frac{1}{3}$ at order $n=2$ implies that $SM$ is an unstable saddle rest point but not an unstable source. From this we conclude that not all solutions in $\mathcal{F}$ lie in the unstable manifold of $SM$, even though they decay time asymptotically to $M$ as $\bar{t}\to\infty$. Since $SM$ is a saddle rest point, backward time trajectories in $\mathcal{F}$ starting near $SM$ will not typically tend to $SM$, but rather, indicative of the standard phase portrait of a saddle rest point, will generically follow the backward time trajectory of the stable manifold at $SM$. This implies the Big Bang is self-similar like the $k=0$ Friedmann spacetime at order $n=1$ but generically not self-similar at higher orders, as recorded in Theorem \ref{T9}. We conclude that the time evolution of perturbations of $SM$ becomes indistinguishable from $k<0$ Friedmann spacetimes at late times after the Big Bang, agrees exactly with the same $k<0$ Friedmann spacetime in the leading order phase portrait, including the limits $\bar{t}\to0$ and $\bar{t}\to\infty$ (for each fixed $\bar{r}>0$), but introduce anomalous accelerations away from $k<0$ Friedmann spacetimes at intermediate times, starting at order $n=2$. This provides a rigorous mathematical framework and mechanism for determining and explaining the source of the corrections to redshift vs luminosity computed numerically in \cite{STV2017}, that is, created by the instability of $SM$.

Note that solutions which follow the stable manifold of $SM$ into the rest point $SM$ in forward time, will evolve more and more like $SM$, but in backward time will have a velocity and density which blow up in the limit $\bar{t}\to0$. Note also that this solution can be matched to a $p=\frac{1}{3}\rho$ pure radiation solution at any positive time (see \cite{STV2017}) and by backward evolution from there we see that there are always big bang solutions of pure radiation which generate all of these $p=0$ solutions from some time onward. The authors find it interesting that the presence of the negative eigenvalue at $SM$ at order $n=2$ therefore suggests a mechanism for inflation within Einstein's theory of gravity without incorporating the cosmological constant, either for inflation, or to model a dark energy in the subsequent dynamics.

The new parameter $b$ associated with $(\lambda_{A2},\boldsymbol{R}_{A2})$ naturally introduces accelerations away from Friedmann spacetimes at order $\xi^2$ in $w$, and hence order $\xi^3$ in the velocity $v$. These mimic the effects of a cosmological constant at third order in redshift factor $\rm{z}$ vs luminosity distance $d_\ell$, as measured from the center (see \cite{STV2017} and Section \ref{S2.9}). This is the order of the discrepancy between the prediction of Friedmann spacetimes with a cosmological constant and Friedmann spacetimes without one. According to Figure \ref{Figure3}, at late times after the Big Bang we should expect to observe spacetimes close to $k<0$ Friedmann spacetimes, but not $k=0$. Moreover, perturbations from $k<0$ Friedmann spacetimes, including perturbations of $SM$ in $\mathcal{F}$ at early times after the Big Bang, diverge from $k<0$ Friedmann spacetimes at intermediate times before they decay back to $k<0$ Friedmann spacetimes at late times. Regarding the intermediate times, the new free parameter $b$ associated with the unstable manifold of the $k=0$ Friedmann spacetime ($SM$) is not present in pure $k<0$ Friedmann spacetimes, and this effect appears to mimic the effects of a cosmological constant at the order (third order in redshift factor looking out from the center) at which the predictions of a cosmological constant diverge from the predictions of the Friedmann spacetimes without one. Said differently, this theory identifies a one parameter family of corrections to Friedmann spacetimes at order $n=2$, with further corrections to Friedmann spacetimes determined by the positive eigenvalues of $SM$ at higher orders, with higher orders corresponding to smaller corrections near the center.

More generally, it was proven in \cite{STV2017} that the order in redshift factor in the relation between redshift and luminosity looking out from the center of a spherically symmetric spacetime is at the same order as $\xi$ in our theory here. The Hubble constant and the quadratic correction to redshift vs luminosity is determined at order $n=1$ from $v=w_1\xi$ and $z_2\xi^2$ respectively, and hence $w_2\xi^3$ determines the third order correction in red-shift factor with $z_4\xi^4$ determining the fourth order term. Since one requires values of $w_2$ and $z_4$ to determine whether a solution trajectory lies in $\mathcal{F}'$, it follows that the fourth order correction to red-shift vs luminosity would be required to determine whether or not a cosmology lies in the unstable manifold of $SM$, that is, to determine whether the Big Bang is self-similar like $SM$ at all orders, or whether it diverges from self-similarity at order $n=2$. We conclude that the family $\mathcal{F}$ extends the $k<0$ Friedmann spacetimes to a stable family of spacetimes, closed under small perturbation, which characterizes the instability of the critical ($k=0$) and underdense ($k<0$) Friedmann spacetimes to smooth radial perturbations.

\subsection{The STV ODE Phase Portrait of Order $n\geq2$}\label{S2.8}

The nested property of the STV ODE in (\ref{nxnSystem1}) implies that lower order eigenvalues of $SM$ persist to higher orders. We prove that two distinct additional eigenvalues always emerge at $SM$ in going from the STV ODE of order $n-1$ to the STV ODE of order $n$ for all $n\geq2$. These are given by (\ref{EigenvaluesSM}). From this, we conclude that both eigenvalues $\lambda_{An}$ and $\lambda_{Bn}$ are positive except at orders $n=1$ and $n=2$, when only $\lambda_{A1}$ and $\lambda_{A2}$ are positive. We argued previously that $\lambda_{B1}=-1$ is eliminated by changing to time since the Big Bang, an assumption equivalent to assuming a solution agrees with a Friedmann spacetime at leading order $n=1$. At order $n=2$, $\lambda_{A2}=\frac{4}{3}>0$ and $\lambda_{B2}=-\frac{1}{3}<0$. A calculation also shows that at second order, the $k<0$ Friedmann spacetimes continue to lie exclusively on the trajectory associated with the leading order eigenvalue $\lambda_{A1}=\frac{2}{3}$. Recall that assuming time since the Big Bang eliminates the leading order negative eigenvalue by transforming the solution space to solutions which agree at order $n=1$ with either $SM$ or a trajectory in the unstable manifold of $SM$. The existence of the negative eigenvalue $\lambda_{B2}$ implies that $SM$ is an unstable saddle rest point with a one-dimensional stable manifold and a codimension one unstable manifold at each order $n\geq2$. Also recall that we let $\mathcal{F}_n'\subset\mathcal{F}_n$ denote the subset of solutions with trajectories in the unstable manifold of $SM$ identified as an $(n-1)$-dimensional space of trajectories taking $SM$ to $M$ in the phase portrait of the STV ODE of order $n\geq2$. The appearance of one positive and one negative eigenvalue at order $n=2$, and only positive eigenvalues at higher orders, immediately implies that a solution in $\mathcal{F}$ lies in the unstable manifold of $SM$ at all orders $n\geq1$ if and only if it lies in the unstable manifold of $SM$ at order $n=2$. This is stated precisely in the following theorem.

\begin{theorem}\label{T8}
    The unstable manifold $\mathcal{F}_n'\subset\mathcal{F}_n$ of $SM$ forms a codimension one set of trajectories in the STV ODE at each order $n\geq2$ and a trajectory lies in $\mathcal{F}_n'$ at every order of the STV ODE if and only if it lies in $\mathcal{F}_2'$. Moreover, trajectories in $\mathcal{F}'$ take $SM$ to $M$ at all orders of the STV ODE, agree with a $k<0$ Friedmann spacetime at order $n=1$ in the limits $\bar{t}\to0$ and $\bar{t}\to\infty$ (for each fixed $\bar{r}>0$) but are generically distinct from, and hence accelerate away from, $k<0$ Friedmann spacetimes at intermediate times in the phase portrait of the STV ODE at every order $n\geq2$.
\end{theorem}

It follows from the theory of non-degenerate hyperbolic rest points that all solution trajectories in $\mathcal{F}_n'$ emerge tangent to the eigendirection associated with the smallest positive eigenvalue at $SM$. Of course at order $n=1$ the leading order eigenvalue associated with the solution trajectory of the $k<0$ Friedmann spacetimes is the smallest eigenvalue and this remains the smallest positive eigenvalue at $n=2$. However, an eigenvalue smaller than this emerges at order $n=3$ namely, $\lambda_{B3}=\frac{1}{3}$. Thus solutions in $\mathcal{F}_n'$ generically emerge tangent to the leading order eigendirection of the $k<0$ Friedmann spacetimes only up to order $n=2$, but all solutions in $\mathcal{F}_n'$ which have a component of $\lambda_{B3}$ will emerge from $SM$ tangent to its eigenvector $\boldsymbol{R}_{B3}$, which is not tangent to the $k<0$ Friedmann direction $\boldsymbol{R}_{A1}$, corresponding to the leading order eigenvalue $\lambda_{A1}=\frac{2}{3}$ at $SM$. From this we establish that although all the solutions in $\mathcal{F}_n$ tend to rest point $M$ as $\bar{t}\to\infty$, solutions in $\mathcal{F}_n\setminus\mathcal{F}_n'$, that is, those that miss $SM$ in backward time, follow the backward stable manifold of $SM$, consistent with the standard phase portrait picture of a non-degenerate saddle rest point. Since the only negative eigenvalue of $SM$ enters at order $n=2$, we can conclude that any solution that lies in the unstable manifold of $SM$ in the phase portrait of the STV ODE at order $n=2$ also lies in the unstable manifold of $SM$ at all higher orders $n\geq3$. In this sense, the unstable manifold of $SM$ is characterized at order $n=2$. Note that all solutions of the STV ODE in $\mathcal{F}'$ take $SM$ to $M$ in the phase portrait of the STV ODE at all orders, and hence are bounded for all time $0\leq\bar{t}\leq\infty$. Trajectories not in the unstable manifold of $SM$ miss $SM$ in backwards time in every STV ODE of order $n\geq2$ and tend in backward time instead to the stable manifold associated with the unique negative eigenvalue, that is, a single trajectory. Thus $\mathcal{F}$, which consists of the domain of attraction of $M$ at every order, contains trajectories which do not emanate from $SM$, and hence correspond to a big bang at $\bar{t}=0$ which is qualitatively different from the self-similar blow-up $\bar{t}\to0$ at $SM$, and hence qualitatively different from the Big Bang observed in Friedmann spacetimes. An immediate conclusion of this analysis is a rigorous characterization of the self-similar nature of the Big Bang in general spherically symmetric smooth solutions to the Einstein--Euler equations when $p=0$.

\begin{theorem}\label{T9}
    When time is taken to be time since the Big Bang, solutions in $\mathcal{F}$ always exhibit self-similar blow-up in the $n=1$ phase portrait of the STV ODE but will generically exhibit non-self-similar blow-up at all higher orders $n\geq2$.
\end{theorem}

\subsection{Cosmological Interpretation}\label{S2.9}

In our 2017 RSPA paper \cite{STV2017}, we explored the possibility that a self-similar wave at the end of the Radiation Dominated Epoch (RDE) of the Big Bang may have been the genesis for the anomalous acceleration observed in the supernova data at present time. The existence of a one parameter family of self-similar solutions which perturb the $k=0$ Friedmann spacetime with equation of state $p=\frac{1}{3}\rho$ led the authors to propose such waves as natural candidates for local time asymptotic wave patterns which may plausibly emerge by the end of the RDE. We conjectured, based on ideas from the mathematical theory of shock waves, that the enormous pressure and modulus of genuine nonlinearity present when $p=\frac{1}{3}\rho$ might reasonably imply decay to a local non-interacting wave pattern by the end of the RDE. Authors then derived the STV PDE for the purpose of evolving these self-similar solutions forward in time from the end of the RDE up through the Matter Dominated Epoch (MDE) to present time, in order to compare the resulting corrections with the effects of dark energy. To make the comparison, we derived formulas for the coefficients in an expansion of redshift vs luminosity distance, as measured by an observer at the center of such an underdensity at present time. That is, writing
\begin{align*}
    H_0d_\ell = \rm{z} + Q\,\rm{z}^2 + C\,\rm{z}^3 + O_{\rm{z}\to0}(\rm{z}^4)
\end{align*}
where $\rm{z}$ is the standard redshift factor and $d_\ell$ the luminosity distance \cite{GH2007,PK2006,STV2017}, we showed that the $n=1$ STV ODE phase portrait determines $H_0$ and $Q$. In fact, restricting to solutions in $\mathcal{F}$, the present value of $w_0$ determines $H_0$ and the present value of $z_2$ determines $Q$.

\begin{figure}[hbt!]
    \centering
    \caption{A vector field phase portrait for the $2\times2$ system.}
    \includegraphics[width=0.8\textwidth]{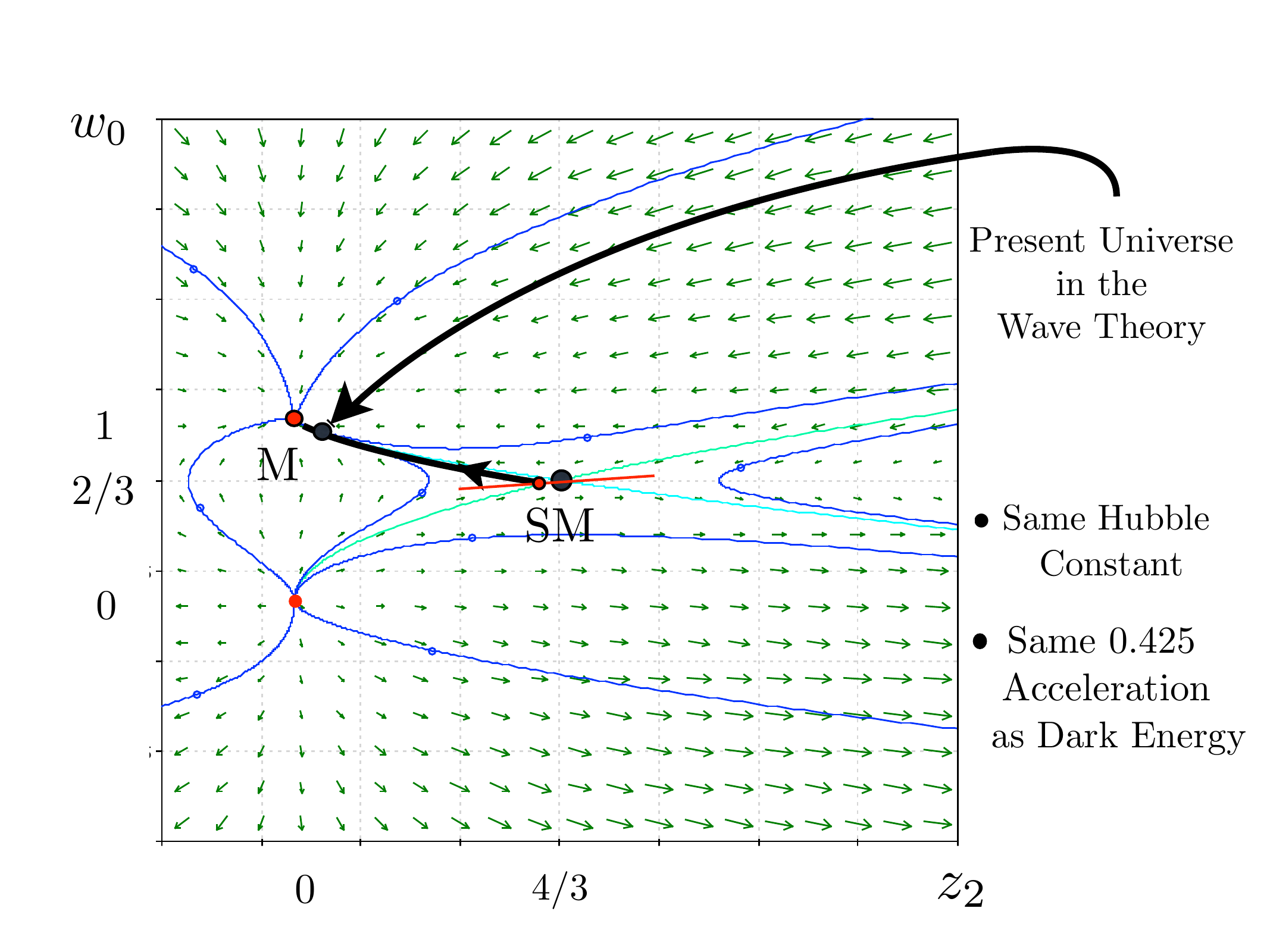}\label{Figure4}
\end{figure}

A calculation using the current model of the $k=0$ Friedmann spacetime with $70\%$ dark energy by virtue of a cosmological constant gives the value $Q=0.425$. By an apparently remarkable coincidence, a calculation in \cite{STV2017} verifies that the $k=0$ Friedmann spacetime with dark energy, as well as the $n=1$ phase portrait of the STV ODE, both produce precisely the same range of possible values of $Q$, namely, both take values within the narrow range $0.25\leq Q\leq0.5$ as $\bar{t}\to\infty$. Figure 4, taken from \cite{STV2017},\footnote{Note that in \cite{STV2017}, the fixed point $SM$ of the $n=1$ phase portrait was centered on $(0,0)$.} shows where the current values of $H_0$ and $Q$ place the Universe at present time according to the $n=1$ STV ODE phase portrait.

Using the STV PDE to evolve self-similar waves at the end of the RDE up to present time, and finding the value of the acceleration parameter associated with those waves that produces the correct values of $H_0$ and $Q$, we made a prediction based on the discrepancy between the value of $C$ implied by $70\%$ dark energy and the value of $C$ determined by the wave theory. The conclusion was:\footnote{The authors note that the third order correction is consistent with an expansion rate that is slowing relative to the predictions of dark energy \cite{A+2025,L+2025}.}
\begin{align*}
    H_0d_\ell &= \rm{z} + 0.425\,\rm{z}^2 - 0.1804\,\rm{z}^3 + O_{\rm{z}\to0}(\rm{z}^4)\qquad \rm{(Dark\ Energy)}\\
    H_0d_\ell &= \rm{z} + 0.425\,\rm{z}^2 + 0.3591\,\rm{z}^3 + O_{\rm{z}\to0}(\rm{z}^4)\qquad \rm{(Wave\ Energy)}
\end{align*}
In fact, the order of $\xi$ in variables $z_n$ and $v_n=\xi w_n$ in a solution of the STV ODE determines the coefficient of redshift factor \rm{z}$^n$ in the expansion of redshift vs luminosity distance. It follows that, in principle, the coefficients in the expansion of redshift vs luminosity up to order $2n$ uniquely determine the initial data for the STV ODE of order $n$ and vice-verse. In particular, this tells us that we would need redshift vs luminosity up to order $n=4$ to determine the solution of the STV ODE up to order $n=2$. Thus for solutions in $\mathcal{F}$, redshift vs luminosity up to order $n=4$ is necessary and sufficient for determining whether the STV ODE solutions lie in the unstable manifold $\mathcal{F}'$ of $SM$ or not, and hence for determining whether the Big Bang is self-similar or not.\footnote{Here we refer to the Big Bang associated with $p=0$, which might have different properties from the Big Bang for $p=\frac{1}{3}\rho$ of the RDE. We note that the procedure of matching metrics at the end of the RDE developed for the simulation in \cite{STV2017} also works to show that any $p=0$ solution can be matched to $p=\frac{1}{3}\rho$ at any given time, and from there run backwards into a big bang associated with the $p=\frac{1}{3}\rho$ of the RDE. In this sense, any solution that comes from a $p=0$ big bang also arises from a $p=\frac{1}{3}\rho$ big bang.} Authors are currently working on extending the stability analysis for $p=0$ solutions of the MDE of the Big Bang, to the case $p=\frac{1}{3}\rho$ associated with the earlier RDE. Our program is to determine whether the instability of the Big Bang itself, rather than local fluctuations from an earlier epoch, might actually be the source of the anomalous acceleration of the galaxies, not dark energy. Indeed, a $k<0$ Friedmann spacetime emerges from the unstable manifold of $SM$ at the Big Bang on par with other solutions which accelerate away from Friedmann spacetimes. Of course, this would require the Milky Way galaxy to be sufficiently centered on that part of the acceleration which is present over and above a background $k<0$ Friedmann spacetime in order to account for the apparent uniformity of the galaxies in opposite directions of observation.

\section{Conclusions}\label{S3}

The discovery of a positive and negative eigenvalue at $SM$ in the phase portrait of the STV ODE at order $n=2$, both different from the leading order eigenvalues associated with $k<0$ Friedmann spacetimes, implies two things: The positive eigenvalue produces a free parameter in solutions associated with accelerations away from $k<0$ Friedmann spacetimes within the unstable manifold $\mathcal{F}'$ of $SM$, and the negative eigenvalue implies the existence of solutions in $\mathcal{F}$ which are arbitrarily close to $SM$, lie outside the unstable manifold $\mathcal{F}'$, but also accelerate away from $k<0$ Friedmann spacetimes before they decay back to $k<0$ Friedmann spacetimes along the unstable manifold $\mathcal{F}'$ of $SM$. The presence of only positive eigenvalues at $SM$ at all orders $n>2$ implies the existence of solutions in which, like $k<0$ Friedmann spacetimes, emerge from $SM$ from within the unstable manifold $SM$, agree with Friedmann spacetimes at order $n=1$, but generically produce accelerations away from Friedmann spacetimes at all orders $n\geq2$, before they decay back to $k<0$ Friedmann spacetimes as $\bar{t}\to\infty$ (for each fixed $\bar{r}>0$). Generic solutions in $\mathcal{F}$ are not in $\mathcal{F}'$, do decay to $k<0$ Friedmann spacetimes along $\mathcal{F}'$ as $\bar{t}\to\infty$, but do not tend to $SM$ in backward time $\bar{t}\to0$, meaning they exhibit a non-self-similar big bang different from Friedmann spacetimes. In particular, this implies solutions could spend an arbitrarily long time arbitrarily close to rest point $SM$ corresponding to the flat $k=0$ Friedmann spacetime, but ultimately evolve away from $SM$ in both forward and backward time. Thus anomalous accelerations away from $k\leq0$ Friedmann spacetimes are not a violation, but a prediction of Einstein's original theory of General Relativity without a cosmological constant. This conclusion does not change with the addition of dark energy because the cosmological constant is negligible relative to the energy density during the early epoch when the instability is triggered. The possibility that the Universe on the largest scale may have evolved from a smooth perturbation of $SM$ which lies within the family $\mathcal{F}$ is compelling, because the instability of $SM$ to perturbations at every order at the Big Bang makes the $k=0$ Friedmann spacetime implausible as a physically observable model, with or without dark energy. Also, in light of our proof that all coefficients in an expansion of luminosity distance in terms of redshift can be accounted for by initial conditions for the STV ODE, the set of possible accelerations induced by the instability is rich enough to mimic the effects of dark energy at all orders, even consistent with a variable acceleration \cite{A+2025,L+2025}. Given the results presented here, the authors find it hard to imagine that a successful explanation for the anomalous acceleration of our Universe is possible without accounting for the instability of the critical Friedmann spacetime at the Big Bang itself.

\end{document}